\documentclass[
 reprint,
 amsmath,amssymb,
 aps,
]{revtex4-2}
\usepackage{graphicx}
\usepackage{dcolumn}
\usepackage{bm}
\usepackage{booktabs}    
\usepackage{dcolumn}     
\newcolumntype{.}{D{.}{.}{4}} 
\usepackage{graphicx}
\usepackage{amsmath}  
\usepackage{hyphenat} 
\makeatletter
\renewenvironment{thebibliography}[1]
 {\section*{\refname}%
  \footnotesize
  \list{\@biblabel{\@arabic\c@enumiv}}%
       {\settowidth\labelwidth{\@biblabel{#1}}%
        \leftmargin\labelwidth
        \advance\leftmargin\labelsep
        \usecounter{enumiv}%
        \let\p@enumiv\@empty
        \renewcommand\theenumiv{\@arabic\c@enumiv}}%
  \sloppy\clubpenalty4000\widowpenalty4000%
  \sfcode`\.\@m}
 {\def\@noitemerr
   {\@latex@warning{Empty `thebibliography' environment}}%
  \endlist}
\makeatother

\usepackage[utf8]{inputenc}
\usepackage{graphicx}
\usepackage{dcolumn}
\usepackage{bm}
\usepackage{float}
\usepackage{amsmath}
\usepackage{graphicx,subfigure,epsfig}
\usepackage{hyperref}
\hypersetup{
    colorlinks=true,
    citecolor=blue,
    linkcolor=blue,
    filecolor=magenta,
    urlcolor=cyan,}
\makeatletter
       
\def\thetable{\@arabic\c@table} 
\makeatother

\begin{document}

\preprint{APS/123-QED}

\title{Testing charged bumblebee black holes through high-frequency quasi-periodic oscillations in X-ray binaries}

\author{Qi-Qi Liang$^{1}$}
\author{De-Jiang Yin$^{1}$}
\author{Zi-Qiang Cai$^{1}$}
\author{Zheng-Wen Long$^{1}$}

\email[Corresponding author: ]{zwlong@gzu.edu.cn}
\affiliation{$^{1}$College of Physics, Guizhou University, Guiyang, 550025, China}

\begin{abstract}
We investigate strong‑field orbital dynamics for static spherically symmetric charged bumblebee black holes featuring spontaneous Lorentz‑violating effects. Analysis of null and timelike circular geodesics yields the photon‑sphere radius, the innermost stable circular orbit (ISCO) radius, and characteristic orbital frequencies for test particles. Within the relativistic precession model for high‑frequency quasi‑periodic oscillations (HFQPOs), observed twin‑peak frequencies are governed by the full set of fundamental parameters $M$, $X=r/M$, $\l_1$, $\l_2$, and $Q_0/M$. Among them, $\l_1$, $\l_2$, and $Q_0/M$ enter frequency expressions only in combined forms, giving rise to intrinsic parameter degeneracy. An effective‑parameter set $\Theta=(M,X,C,\beta)$ is accordingly introduced to characterize these composite contributions. Bayesian MCMC parameter inference is carried out using HFQPO observational data from three black‑hole X‑ray binaries: GRO J1655--40, XTE J1550--564, and GRS 1915+105. The 68\% credible intervals of composite parameter $C$ all contain the Reissner--Nordstr\"om limit $C=1$, revealing no statistically significant net Lorentz‑violating correction under our model assumptions. Distinct triples $(\l_1,\l_2,Q_0/M)$ yield identical QPO predictions, so HFQPO observations alone cannot disentangle Lorentz‑violating and charge‑related effects; precise constraints on these fundamental quantities require additional, more precise observational data.

\end{abstract}

\maketitle
\section{Introduction}
Lorentz symmetry constitutes one of the fundamental spacetime symmetries underlying both general relativity and the Standard Model of particle physics. Nevertheless, in several candidate theories of quantum gravity, Lorentz symmetry may be broken or nontrivially modified at fundamental energy scales, with possible consequences surviving in their low-energy effective descriptions~\cite{Kostelecky:2003fs,Colladay:1998fq,Colladay:1996iz,Horava:2009uw,Carroll:2001ws}. In the case of spontaneous Lorentz violation, the underlying action can remain invariant, while a background field with a nonvanishing vacuum expectation value selects a preferred spacetime direction in the vacuum. Consequently, the vacuum configuration no longer respects the full local Lorentz symmetry, leading to modifications of gravitational dynamics~\cite{Kostelecky:1988zi,Bluhm:2008yt}. Bumblebee gravity provides a representative realization of this mechanism, in which a vector field acquires a nonzero vacuum expectation value and couples nonminimally to spacetime curvature, thereby introducing Lorentz-violating effects into the gravitational field equations~\cite{Kostelecky:1989jw,Bertolami:2005bh}. Black-hole solutions in this framework therefore provide a concrete setting for investigating how spontaneous Lorentz violation manifests itself in the strong-field regime and whether it can lead to observable signatures.

Within bumblebee gravity, Casana \textit{et al.} obtained an exact static and spherically symmetric Schwarzschild-like black-hole solution, in which the Lorentz-violating parameter modifies the radial metric component~\cite{Casana:2017jkc}. Subsequent investigations have explored various observational and theoretical aspects of bumblebee black holes, including gravitational lensing~\cite{Ovgun:2018ran}, accretion phenomena~\cite{Yang:2018zef}, finite-distance deflection effects~\cite{Li:2020dln}, black holes with a cosmological constant~\cite{DCarvalho:2021zpf}, rotating extensions~\cite{Jha:2020pvk}, and quasinormal-mode properties~\cite{Oliveira:2021abg}. More general vacuum configurations have further revealed that different bumblebee field structures and coupling conditions can generate distinct classes of black-hole geometries~\cite{Xu:2022frb}. In addition, metric-affine formulations~\cite{Filho:2022yrk}, higher-dimensional and (A)dS extensions~\cite{Uniyal:2022xnq}, slowly rotating perturbations~\cite{Liu:2022dcn}, and stationary axisymmetric solutions~\cite{AraujoFilho:2024ykw} have expanded the phenomenological landscape of bumblebee black holes. These studies indicate that Lorentz-violating modifications can significantly affect strong-field structures and motivate further investigations of their observable consequences.

High-frequency quasi-periodic oscillations (HFQPOs) observed in black-hole X-ray binary systems provide a valuable probe of strong-field orbital dynamics near compact objects~\cite{Remillard:2006fc}. Although the physical origin of HFQPOs remains under debate, several phenomenological models relate their characteristic frequencies to orbital and epicyclic motions of matter in the innermost accretion region. These frequencies are therefore sensitive to the underlying spacetime geometry around the central black hole~\cite{Stella:1998mq,Stella:1997tc,Cadez:2008iv,Kostic:2009hp,Abramowicz:2003xy,Kluzniak:2002bb,Nowak:1996hg,Torok:2010rk,Kotrlova:2020pqy}. Among these models, the relativistic precession model (RPM) provides a direct connection between observed QPO frequencies and the fundamental geodesic frequencies of test-particle motion, making it a useful framework for studying modified black-hole spacetimes~\cite{Stella:1999sj,Motta:2013wwa,Ingram:2014ara,Maselli:2014fca}. Therefore, HFQPO observations offer a potential approach to constrain deviations from standard black-hole geometries in theories with modified gravity.

Recently, QPO observations have been applied to investigate several modified black-hole spacetimes with Lorentz-violating features~\cite{Ahmed:2026hmy,Jumaniyozov:2025dyy,Jumaniyozov:2025wcs,Jumaniyozov:2024eah}. Previous studies have analyzed the orbital frequencies of Einstein--bumblebee black holes and used microquasar HFQPO observations to constrain the corresponding Lorentz-violating parameters~\cite{Wang:2021gtd,Zhang:2025acq,Mustafa:2024mvx}. However, these analyses have mainly focused on simpler bumblebee configurations with fewer independent parameters. The QPO signatures of charged bumblebee black holes arising from generalized vacuum structures, as well as the possible parameter degeneracies among Lorentz-violating and charge contributions, have not been systematically explored.

Liu \textit{et al.}recently constructed a class of exact black-hole solutions in bumblebee gravity by considering generalized vacuum expectation values of the bumblebee field~\cite{Liu:2025oho}. In this framework, two independent nonvanishing components of the vector field introduce two Lorentz-violating parameters, $\l_1$ and $\l_2$. After including the electromagnetic sector, a charged static and spherically symmetric black-hole solution with a Reissner--Nordström-like structure was obtained. The additional Lorentz-violating contributions modify the spacetime geometry and consequently affect the characteristic circular orbits and orbital frequencies of test particles.

In this work, we investigate the strong-field observational signatures of the charged bumblebee black-hole solution constructed by Liu \textit{et al.} through HFQPO observations of X-ray binaries. By constructing the relativistic precession model and combining it with Bayesian parameter estimation, we explore the constraints imposed by QPO observations on this modified spacetime. Rather than directly constraining the original parameters, we find that the Lorentz-violating parameters $\l_1$, $\l_2$, and the charge parameter $Q_0/M$ enter the QPO observables only through specific combinations, leading to intrinsic degeneracies in the original parameter space. Therefore, we introduce an effective parameterization that captures the combinations directly probed by observations and constrain them using the HFQPO data of GRO J1655--40, XTE J1550--564, and GRS 1915+105.

This paper is organized as follows. In Sec.~\ref{sec:2}, we introduce the charged bumblebee black-hole solution and discuss its basic geometric properties. In Sec.~\ref{sec:3}, we investigate the null and timelike geodesic structures and analyze the characteristic circular orbits. In Sec.~\ref{sec:4}, we derive the orbital angular frequency and radial epicyclic frequency of test particles. In Sec.~\ref{sec:5}, we construct the HFQPO model within the relativistic precession framework, compare the theoretical frequencies with observations of three X-ray binary systems, and constrain the effective parameters using the Markov chain Monte Carlo (MCMC) method. Finally, Sec.~\ref{sec:6} summarizes our conclusions. Throughout Secs.~\ref{sec:2} and \ref{sec:3}, we adopt geometrized units with $G=c=1$ and set the black-hole mass scale to $M=1$. Physical frequency units are restored in Secs.~\ref{sec:4} and \ref{sec:5} when comparing theoretical predictions with HFQPO observations.

\label{sec:1}

\section{Charged bumblebee black-hole geometry}
\label{sec:2}

The bumblebee gravitational action considered in this work can be written as~\cite{Casana:2017jkc,Liu:2025oho}

\begin{equation}
\begin{aligned}
S=\int d^{4}x\sqrt{-g}\Bigg[
&\frac{1}{2\kappa}(R-2\Lambda)
\\+\frac{\xi}{2\kappa}B^{\mu}B^{\nu}R_{\mu\nu} 
&-\frac{1}{4}B_{\mu\nu}B^{\mu\nu}
-V(X)
\Bigg]
+\int d^{4}x\sqrt{-g}\,\mathcal{L}_{M}.
\end{aligned}
\label{eq1}
\end{equation}

We consider the charged black-hole solution recently obtained in bumblebee gravity by Liu \textit{et al.}~\cite{Liu:2025oho}. In this framework, the vector field $B_{\mu}$ acquires a nonvanishing vacuum expectation value, thereby spontaneously breaking Lorentz symmetry, while its nonminimal coupling to spacetime curvature introduces the corresponding Lorentz-violating effects into the gravitational sector. In the generalized vacuum configuration adopted for this solution, the vacuum expectation value of the bumblebee field contains both temporal and radial components~\cite{Xu:2022frb},
\begin{equation}
b_{\mu}=\left(\alpha,b_r(r),0,0\right),
\label{eq2}
\end{equation}
with $b_{\mu}b^{\mu}=b^2=\mathrm{const}$. The two Lorentz-violating parameters are defined as
\begin{equation}
\l_1=\xi b^2,\qquad
\l_2=\xi\alpha^2,
\label{eq3}
\end{equation}
where $\xi$ denotes the nonminimal coupling between the bumblebee field and gravity. The parameter $\l_1$ characterizes the Lorentz-violating contribution associated with the invariant norm $b^2$ of the bumblebee vacuum expectation value, whereas $\l_2$ represents the additional contribution arising from the nonvanishing temporal component $b_t=\alpha$. For the spacelike vacuum configuration considered here, $b^2>0$, while $\alpha^2\geq0$. Consequently, both $\l_1$ and $\l_2$ inherit their sign from the same coupling constant $\xi$ and therefore cannot have opposite signs; in particular, $\l_1\l_2\geq0$, with $\l_1\l_2>0$ when both components give nonvanishing Lorentz-violating contributions. Although $\l_1$ and $\l_2$ originate from distinct parts of the bumblebee vacuum configuration and enter the black-hole geometry through different combinations, their physically allowed signs are thus correlated.

\begin{figure*}[t]
\centering

\begin{minipage}{0.33\textwidth}
    \centering
    \begin{minipage}{0.9\linewidth}
        \includegraphics[width=\linewidth]{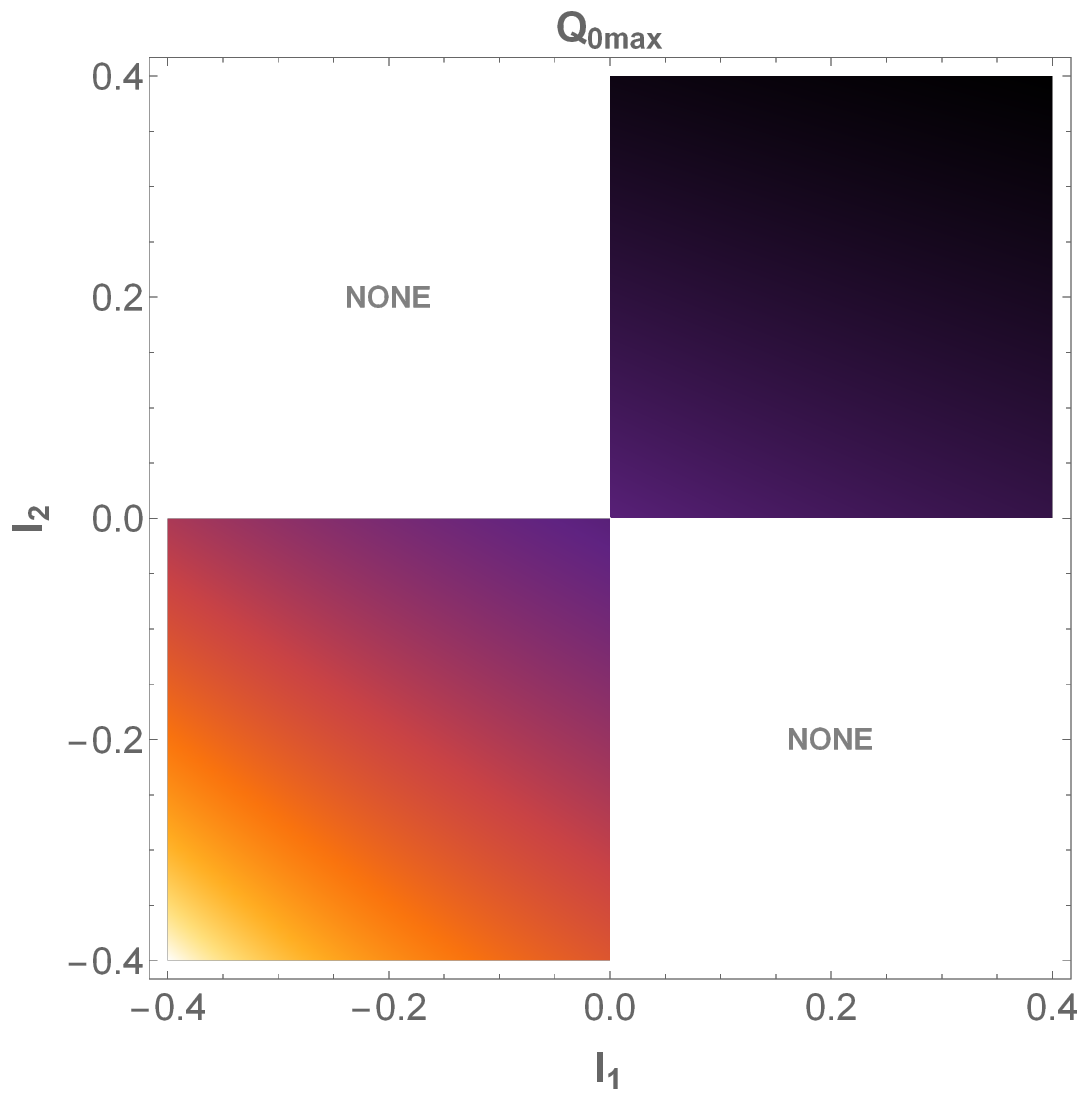}
    \end{minipage}%
    \begin{minipage}{0.08\linewidth}
        \includegraphics[width=\linewidth]{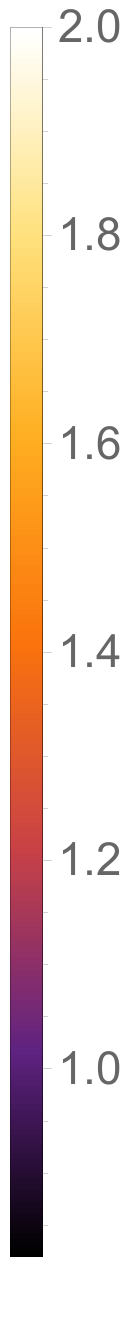}
    \end{minipage}
\end{minipage}%
\begin{minipage}{0.33\textwidth}
    \centering
    \begin{minipage}{0.9\linewidth}
        \includegraphics[width=\linewidth]{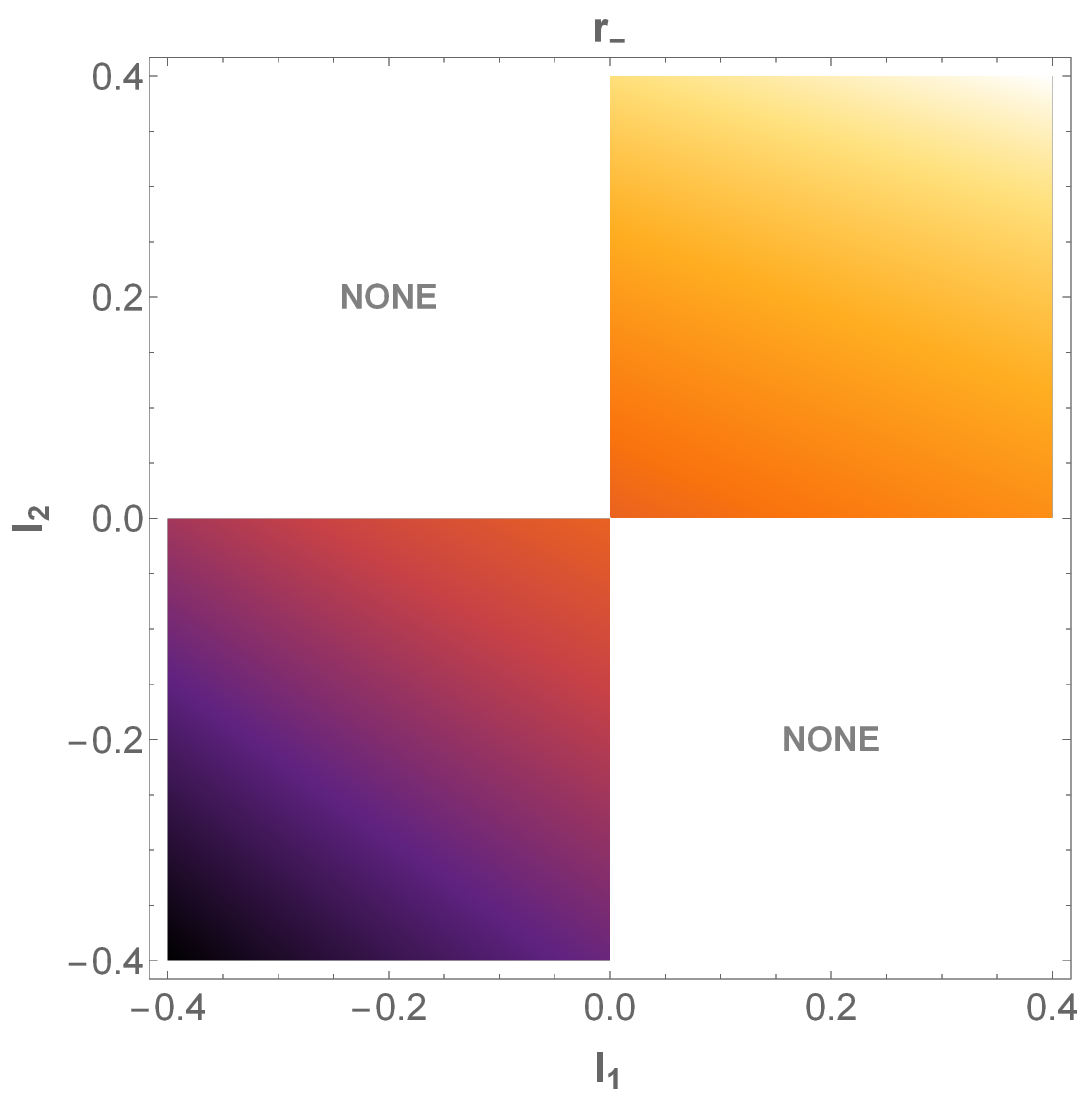}
    \end{minipage}%
    \begin{minipage}{0.08\linewidth}
        \includegraphics[width=\linewidth]{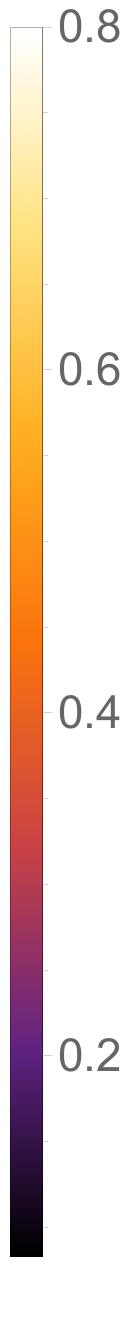}
    \end{minipage}
\end{minipage}%
\begin{minipage}{0.33\textwidth}
    \centering
    \begin{minipage}{0.9\linewidth}
        \includegraphics[width=\linewidth]{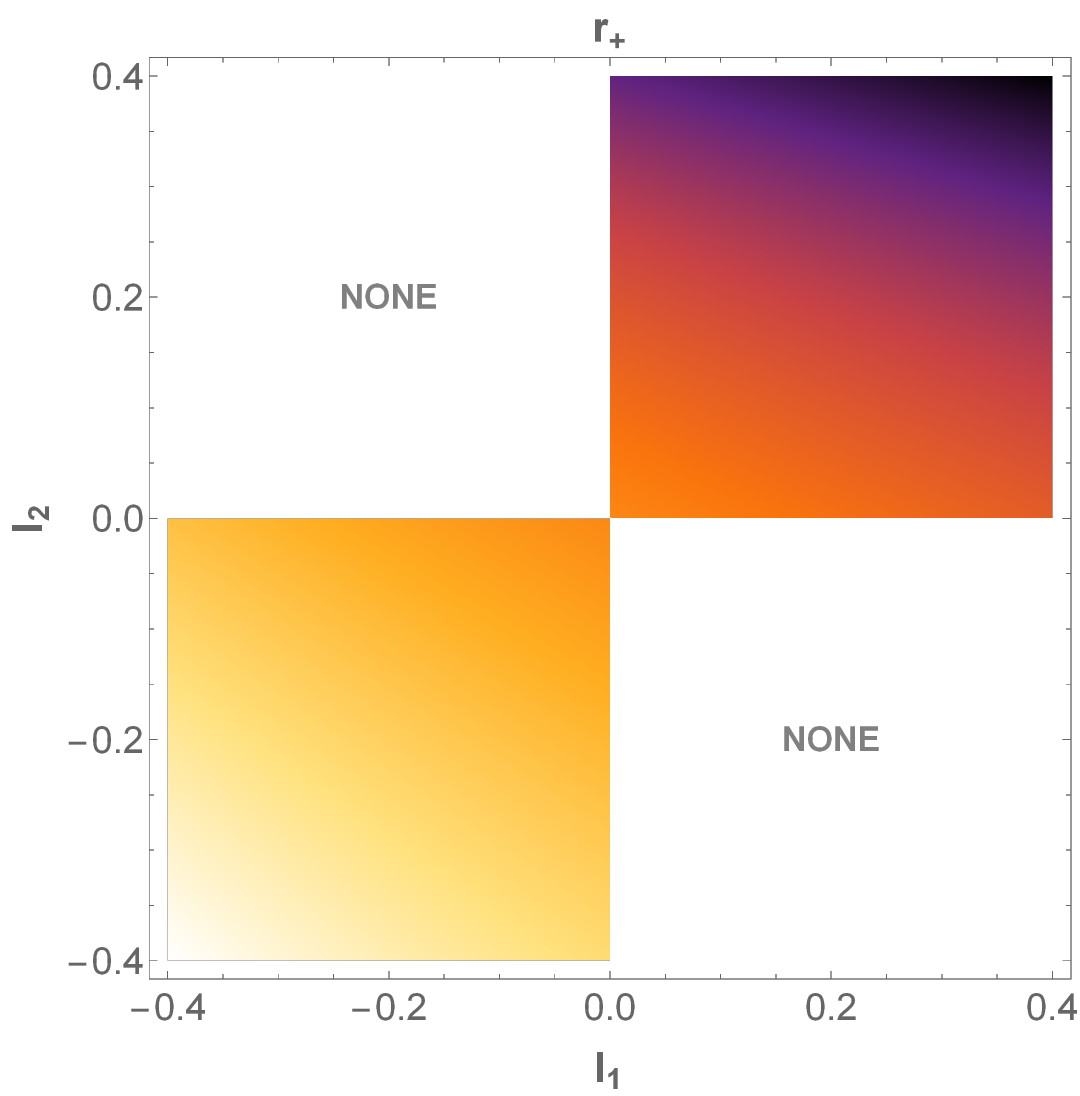}
    \end{minipage}%
    \begin{minipage}{0.08\linewidth}
        \includegraphics[width=\linewidth]{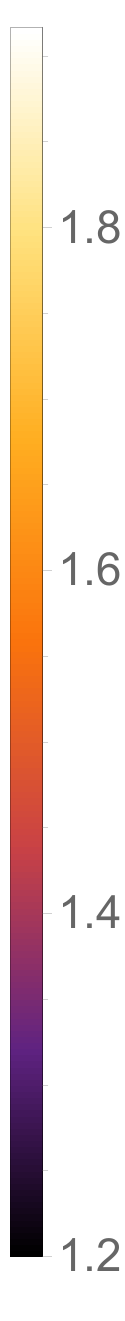}
    \end{minipage}
\end{minipage}
\caption{The influence of Lorentz-violating parameters, with $\l_1\l_2\geq0$, on the charge bound and horizon structure of the charged bumblebee black hole. The left panel displays the maximal allowed charge $Q_{0,\max}$. The middle and right panels show the corresponding inner and outer horizon radii for $M=1$ and $Q_0=0.8$, respectively.}
\label{fig:1}
\end{figure*}

To construct the charged solution, the matter sector is taken to be an electromagnetic field nonminimally coupled to the bumblebee vector field. Following Ref.~\cite{Liu:2024axg}, the corresponding Lagrangian density is written as
\begin{equation}
\mathcal{L}_{M}
=
-\frac{1}{2\kappa}
\left(
F^{\mu\nu}F_{\mu\nu}
+\gamma B^{\mu}B_{\mu}F^{\alpha\beta}F_{\alpha\beta}
\right)
\label{eq4}
\end{equation}
where
\begin{equation}
F_{\mu\nu}
=
\partial_{\mu}A_{\nu}
-
\partial_{\nu}A_{\mu}
\label{eq5}
\end{equation}
\begin{equation}
A_{\mu}=\left(\phi(r),0,0,0\right)
\label{eq6}
\end{equation}
where $F_{\mu\nu}$ is the electromagnetic field-strength tensor and $A_\mu$ denotes the electromagnetic four-potential.$\gamma$ denotes the coupling coefficient between the electromagnetic and bumblebee fields.
After introducing the electromagnetic sector and its nonminimal coupling to the bumblebee field, the field equations admit a static and spherically symmetric charged solution of the form
\begin{equation}
ds^2=-A(r)dt^2+B(r)dr^2+r^2
\left(d\theta^2+\sin^2\theta\,d\phi^2\right)
\label{eq7}
\end{equation}
where
\begin{equation}
A(r)=1-\frac{2M}{r}
+\frac{2(1+\l_1+\l_2)Q_0^2}
{(2+\l_1)r^2}
\label{eq8}
\end{equation}
and
\begin{equation}
B(r)=\frac{1+\l_1+\l_2}{A(r)}
\label{eq9}
\end{equation}

Here $M$ is the mass parameter and $Q_0$ is an integration constant associated with the electromagnetic field. Owing to the nonminimal coupling, note that $Q_0/M$ is the charge parameter appearing in the metric rather than the conserved electric charge $Q$, which is related to it by
\begin{equation}
Q=\frac{2(1+\l_1)}{2+\l_1}Q_0
\label{eq10}
\end{equation}

The solution reduces to the Reissner--Nordstr\"om geometry when $\l_1=\l_2=0$, while setting $\l_2=0$ recovers the corresponding one-parameter charged bumblebee solution. At large radial distances,
\begin{equation}
A(r)\rightarrow1,
\qquad
B(r)\rightarrow1+\l_1+\l_2,
\label{eq11}
\end{equation}
showing that the Lorentz-violating deformation leaves a constant modification in the radial sector of the asymptotic geometry.

The horizon radii are determined by $A(r)=0$ and are given by
\begin{equation}
r_{\pm}
=
M\pm
\sqrt{
M^2-
\frac{2(1+\l_1+\l_2)Q_0^2}
{2+\l_1}
}.
\label{eq12}
\end{equation}
Therefore, a black-hole configuration requires
\begin{equation}
M^2\geq
\frac{2(1+\l_1+\l_2)Q_0^2}
{2+\l_1},
\label{eq13}
\end{equation}
with equality corresponding to the extremal case.For the following analysis,  
Under this normalization, Eq.~(\ref{eq13}) reduces to the constraint on the charge parameter,

\begin{equation}
\frac{|Q_0|}{M}\leq\frac{Q_{0,\max}}{M}=\sqrt{\frac{2+\l_1}{2(1+\l_1+\l_2)}}.
\label{eq14}
\end{equation}
The equality corresponds to the extremal black-hole configuration, where the inner and outer horizons coincide.The dependence of $Q_{0,\max}$ on the Lorentz-violating parameters is shown in the left panel of Fig.~\ref{fig:1}. 

The Lorentz-violating parameters are constrained by the asymptotic behavior of the radial metric component,$B(r)=\frac{1+\l_1+\l_2}{A(r)}$.
Therefore, we consider the following parameter range,
\begin{equation}
-0.4\leq\l_1,\l_2\leq0.4
\label{eq15}
\end{equation}
to investigate how the Lorentz-violating parameters affect the inner and outer horizon structures of the charged bumblebee black hole.The left panel of Fig.~\ref{fig:1} shows the dependence of $Q_{0,\max}$ on the Lorentz-violating parameters. 
Within the parameter range $-0.4\leq\l_1,\l_2\leq0.4$, the maximal allowed charge parameter varies approximately in the range $0.8\sim2.0$. 
As the Lorentz-violating parameters increase, the maximal allowed charge parameter generally decreases. 
To ensure that the black-hole existence condition is satisfied throughout the considered parameter space, we choose $Q_0=0.8$, which remains below the minimum value of $Q_{0,\max}$ in this region. 
Substituting this charge parameter into Eq.~(\ref{eq12}), the corresponding inner and outer horizon radii are displayed in the middle and right panels of Fig.~\ref{fig:1}, respectively. 
It can be seen that increasing $\l_1$ and $\l_2$ gradually enlarges the inner horizon radius while reducing the outer horizon radius. 
This indicates that the Lorentz-violating effects modify the horizon structure of the charged bumblebee black hole and drive the system toward a more extremal configuration.

\section{Null and timelike geodesics}
\label{sec:3}
The spacetime geometry described by Eqs.~(\ref{eq8}) and (\ref{eq9}) determines the motion of photons and massive particles around the charged bumblebee black hole. Since circular orbits provide the basis for studying strong-field dynamics and constructing HFQPO frequencies, we investigate the null and timelike geodesic structures in this section. We first consider the null geodesics.For the static and spherically symmetric spacetime described by Eq.~(\ref{eq7}), the photon motion can be restricted to the equatorial plane, $\theta=\pi/2$, without loss of generality. The corresponding geodesic Lagrangian is defined as

\begin{equation}
\begin{aligned}
\mathcal{L}
&=
\frac{1}{2}g_{\mu\nu}
\frac{dx^{\mu}}{d\lambda}
\frac{dx^{\nu}}{d\lambda}
\\
&=
\frac{1}{2}
\left(
-A(r)\dot{t}^{\,2}
+B(r)\dot{r}^{\,2}
+r^{2}\dot{\theta}^{\,2}
+r^{2}\sin^{2}\theta\,\dot{\phi}^{\,2}
\right)
\end{aligned}
\label{eq16}
\end{equation}
where $\lambda$ denotes the affine parameter along the photon trajectory.
Due to the time-translation and rotational symmetries of the spacetime, the corresponding conserved quantities associated with the Killing vectors $\partial_t$ and $\partial_\phi$ are

\begin{equation}
E=A(r)\dot{t}\qquad
L=r^{2}\dot{\phi}
\label{eq17}
\end{equation}
where $E$ and $L$ represent the conserved energy and angular momentum of the photon, respectively.Using the null condition, the radial equation becomes

\begin{equation}
\dot r^2=
\frac{1}{1+\l_1+\l_2}
\left(
E^2-\frac{A(r)}{r^2}L^2
\right)
\label{eq18}
\end{equation}

Introducing the effective potential
$V_{\rm eff}^{\rm null}=\frac{L^2A(r)}{r^2}$ the photon circular orbit satisfies

\begin{equation}
V_{\rm eff}^{\rm null}(r_{\rm ph})=E^2
\qquad
\left.
\frac{dV_{\rm eff}^{\rm null}}{dr}
\right|_{r=r_{\rm ph}}=0
\label{eq19}
\end{equation}

Substituting the metric function in Eq.~(\ref{eq8}) into the photon sphere condition, the radius of the outer photon circular orbit is obtained as

\begin{equation}
r_{\rm ph}=\frac{3M+\sqrt{9M^2-\frac{16(1+\l_1+\l_2)Q_0^2}{2+\l_1}}}{2}
\label{eq20}
\end{equation}

The critical impact parameter associated with the photon sphere:

\begin{equation}
b_c=\frac{r_{\rm ph}}
{\sqrt{1-\frac{2M}{r_{\rm ph}}
+\frac{2(1+\l_1+\l_2)Q_0^2}
{(2+\l_1)r_{\rm ph}^{2}}}}.
\label{eq21}
\end{equation}

For massive particles, the normalization condition is $g_{\mu\nu}\dot{x}^{\mu}\dot{x}^{\nu}=-1$.
Using the conserved quantities introduced above, the radial motion is expressed as

\begin{equation}
\dot r^2=
\frac{1}{1+\l_1+\l_2}
\left[
E^2-
A(r)
\left(1+\frac{L^2}{r^2}\right)
\right]
\label{eq22}
\end{equation}

The effective potential is therefore $V_{\rm eff}^{\rm timelike}=A(r)\left(1+\frac{L^2}{r^2}\right).$ Circular motion of massive particles corresponds to a constant radial coordinate, for which the radial velocity and radial acceleration vanish simultaneously. Therefore, the energy and angular momentum of circular orbits are determined by

\begin{equation}
V_{\rm eff}^{\rm timelike}=E^2
\qquad
\frac{dV_{\rm eff}^{\rm timelike}}{dr}=0
\label{eq23}
\end{equation}

Solving these equations, the conserved energy and angular momentum of circular orbits are obtained as

\begin{equation}
E^2=
\frac{2A(r)^2}{2A(r)-rA(r)'}=\frac{\left(r^2-2Mr+\dfrac{2(1+\l_1+\l_2)}{2+\l_1}Q_0^2\right)^2}{r^2\left(r^2-3Mr+\dfrac{4(1+\l_1+\l_2)}{2+\l_1}Q_0^2\right)}
\label{eq24}
\end{equation}

\begin{equation}
L^2=\frac{r^3A(r)'}{2A(r)-rA(r)'} =\frac{r^2\left(Mr-\dfrac{2(1+\l_1+\l_2)}{2+\l_1}Q_0^2\right)}{r^2-3Mr+\dfrac{4(1+\l_1+\l_2)}{2+\l_1}Q_0^2}.
\label{eq25}
\end{equation}

The stability of circular orbits is determined by the second derivative of the effective potential. The transition between stable and unstable circular orbits defines the ISCO, which satisfies

\begin{equation}
\left.
\frac{d^2V_{\rm eff}^{\rm timelike}}{dr^2}\right|_{r=r_{\rm ISCO}}=0 
\label{eq26}
\end{equation}

For the metric function given in Eq.~(\ref{eq8}), this condition reduces to

\begin{equation}
\begin{aligned}
r_{\rm ISCO}^{3}-6M r_{\rm ISCO}^{2}+\frac{18(1+\l_1+\l_2)Q_0^2}{2+\l_1}
r_{\rm ISCO}\\-\frac{16(1+\l_1+\l_2)^2Q_0^4}{M(2+\l_1)^2}=0 .
\label{eq27}
\end{aligned}
\end{equation}

\begin{figure*}[t]
\centering
\begin{minipage}{0.33\textwidth}
    \centering
    \begin{minipage}{0.9\linewidth}
        \includegraphics[width=\linewidth]{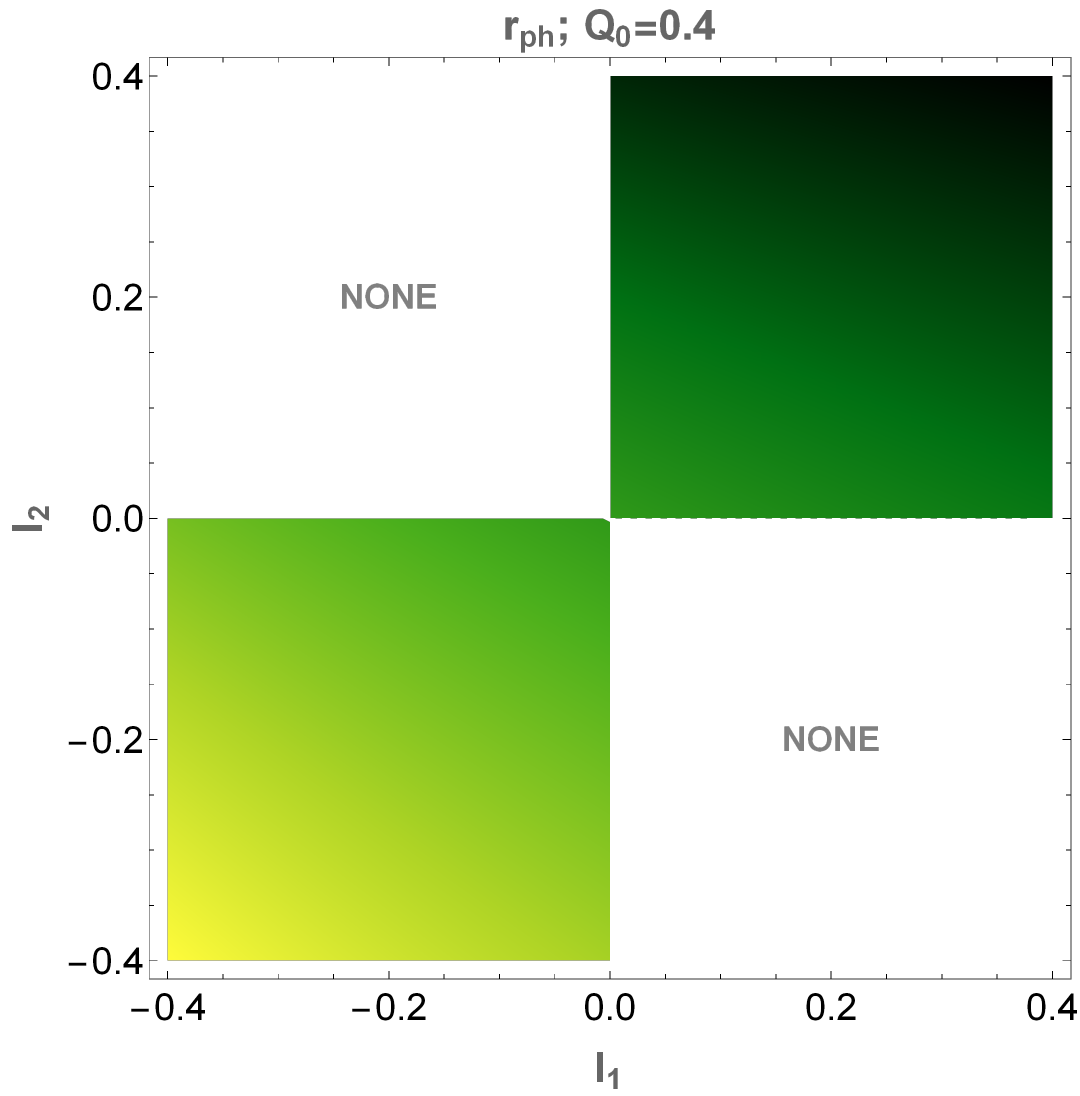}
    \end{minipage}%
    \begin{minipage}{0.09\linewidth}
        \includegraphics[width=\linewidth]{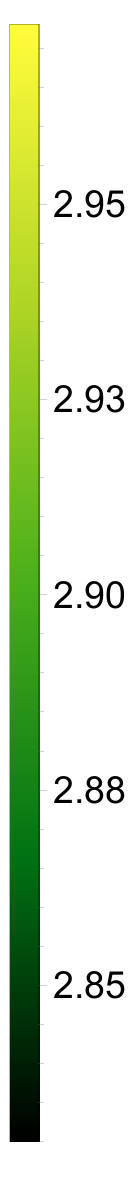}
    \end{minipage}
\end{minipage}%
\begin{minipage}{0.33\textwidth}
    \centering
    \begin{minipage}{0.9\linewidth}
        \includegraphics[width=\linewidth]{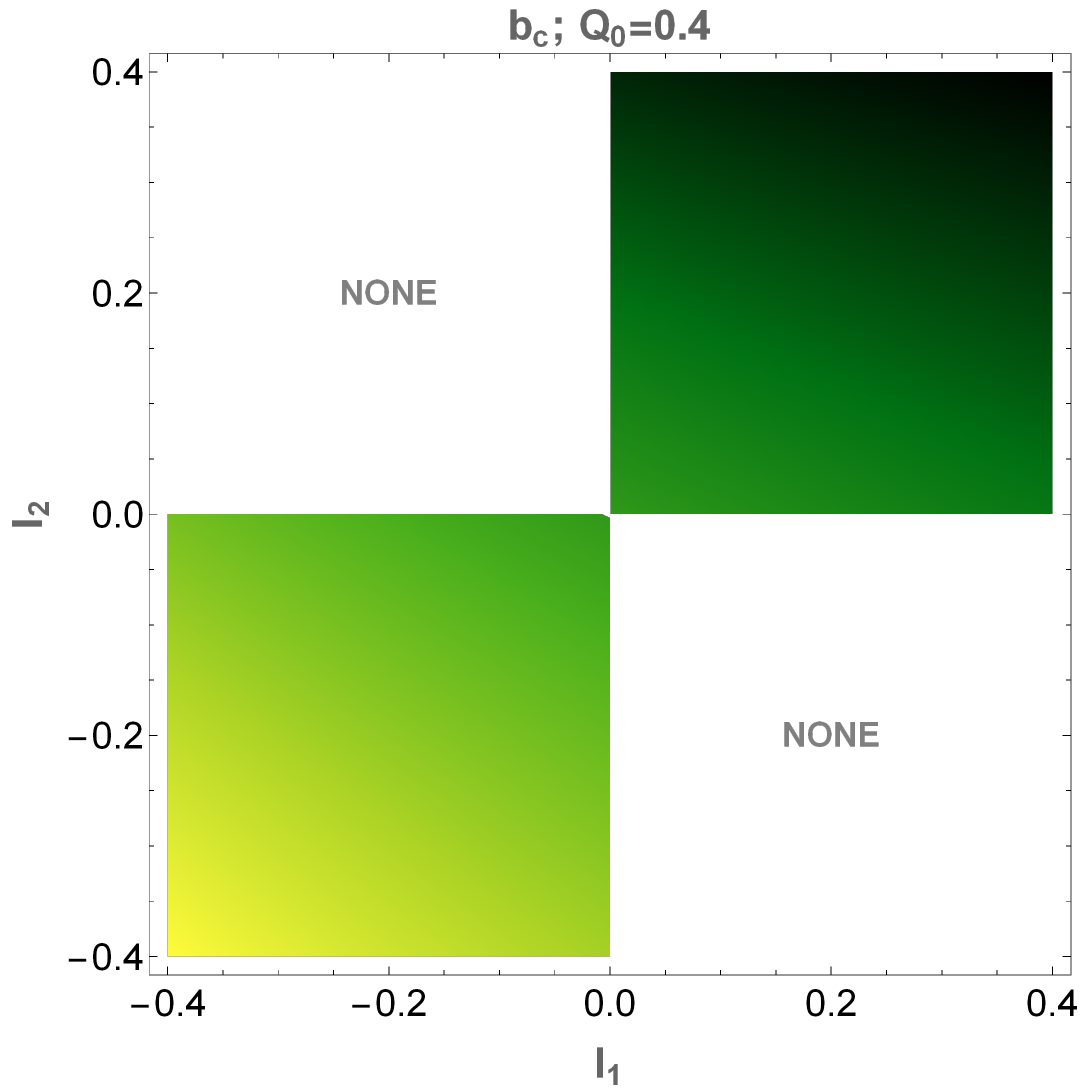}
    \end{minipage}%
    \begin{minipage}{0.09\linewidth}
        \includegraphics[width=\linewidth]{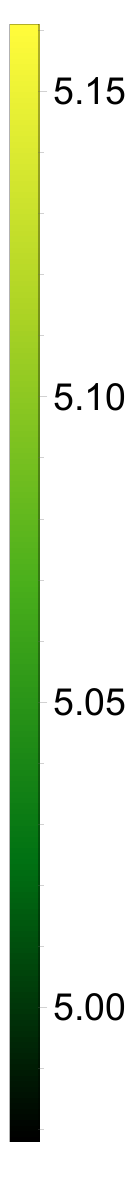}
    \end{minipage}
\end{minipage}%
\begin{minipage}{0.33\textwidth}
    \centering
    \begin{minipage}{0.9\linewidth}
        \includegraphics[width=\linewidth]{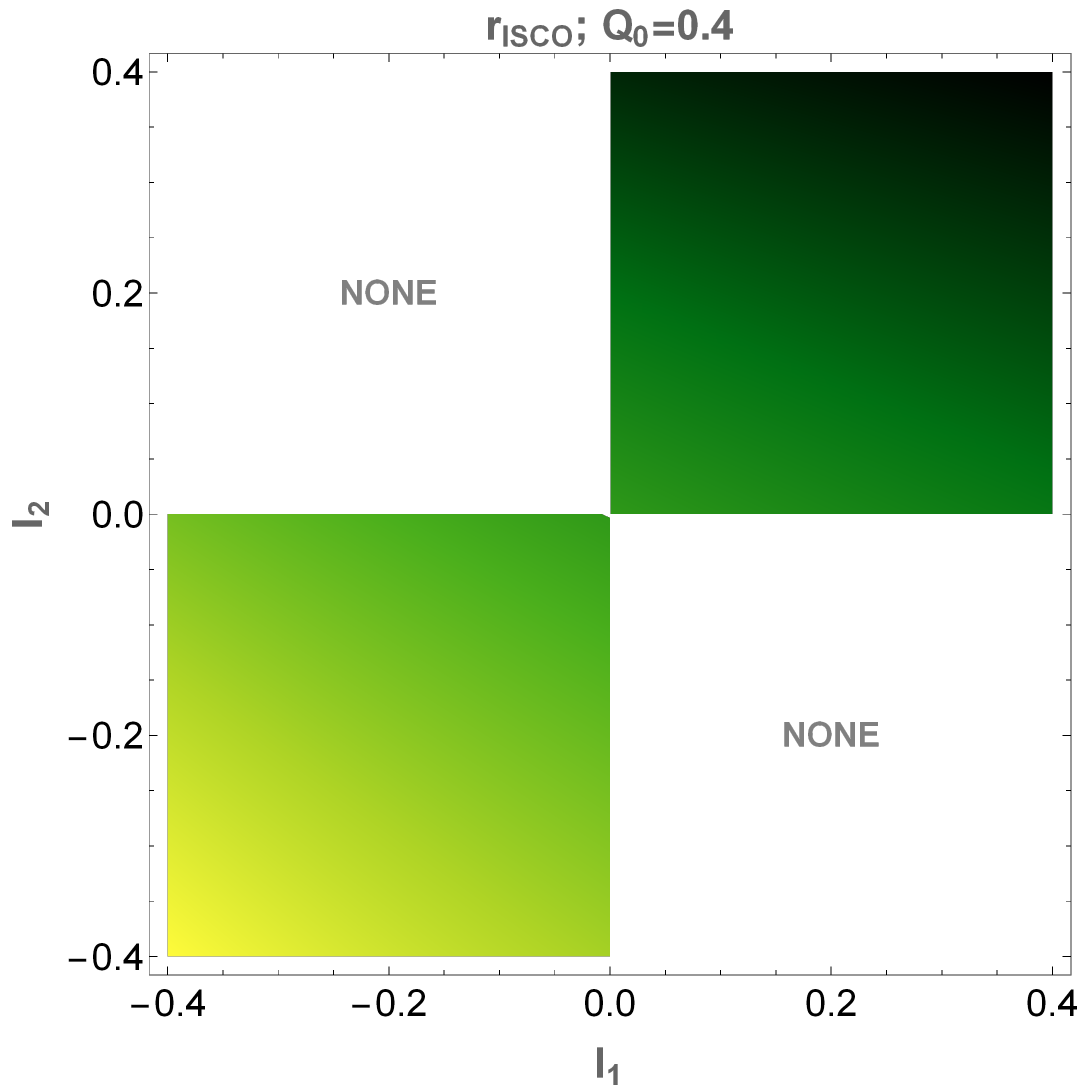}
    \end{minipage}%
    \begin{minipage}{0.09\linewidth}
        \includegraphics[width=\linewidth]{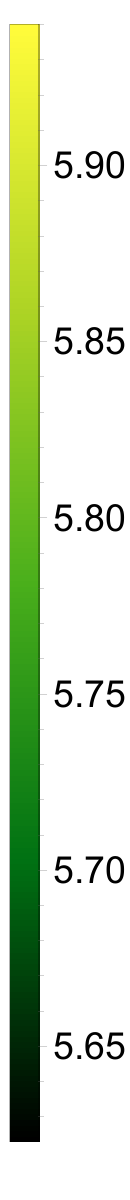}
    \end{minipage}
\end{minipage}

\vspace{0.2cm}

\begin{minipage}{0.33\textwidth}
    \centering
    \begin{minipage}{0.9\linewidth}
        \includegraphics[width=\linewidth]{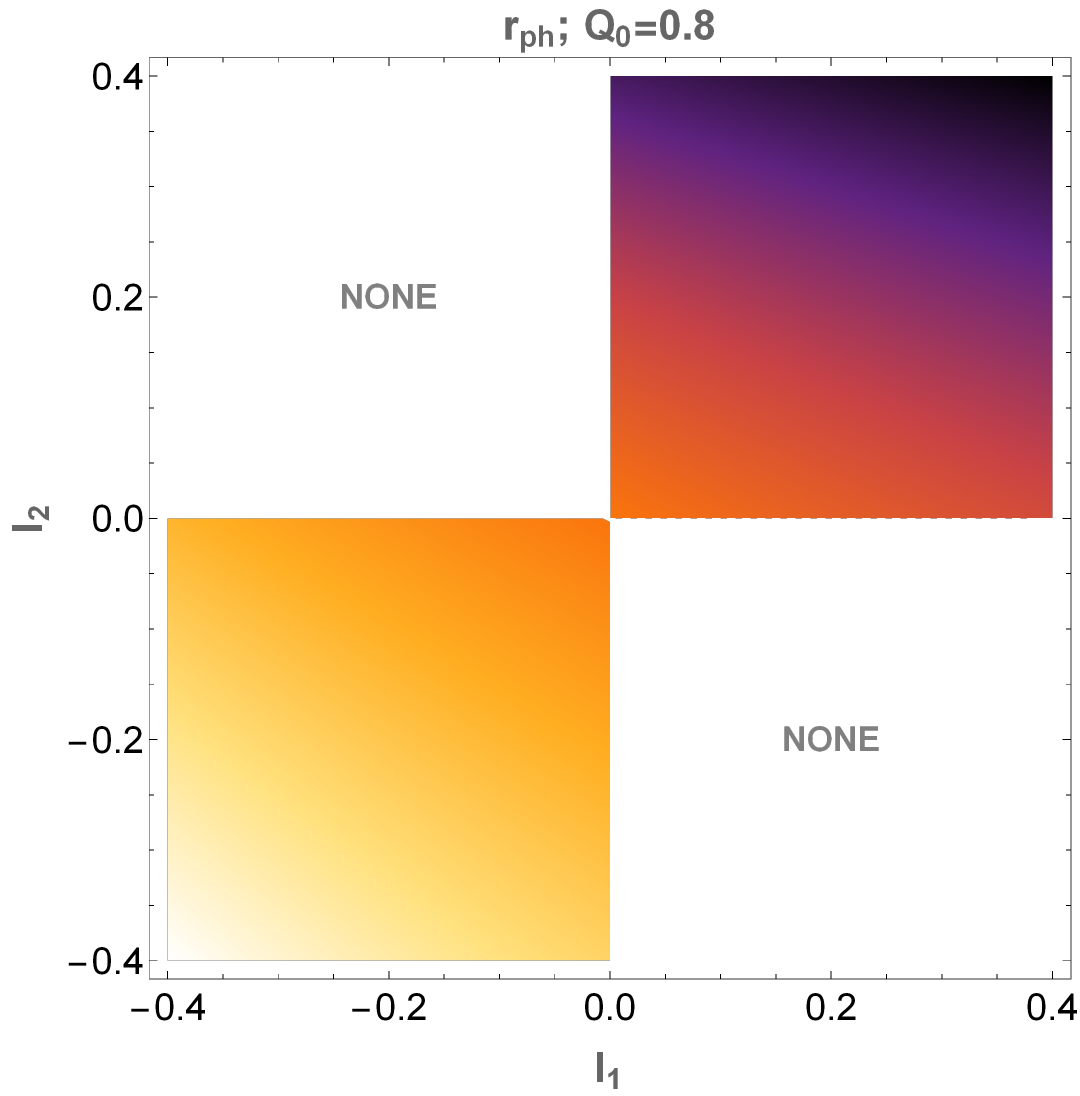}
    \end{minipage}%
    \begin{minipage}{0.09\linewidth}
        \includegraphics[width=\linewidth]{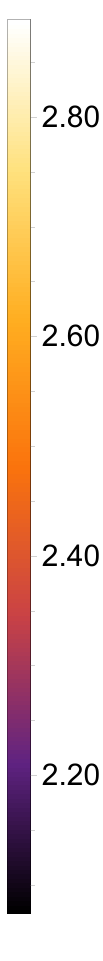}
    \end{minipage}
\end{minipage}%
\begin{minipage}{0.33\textwidth}
    \centering
    \begin{minipage}{0.9\linewidth}
        \includegraphics[width=\linewidth]{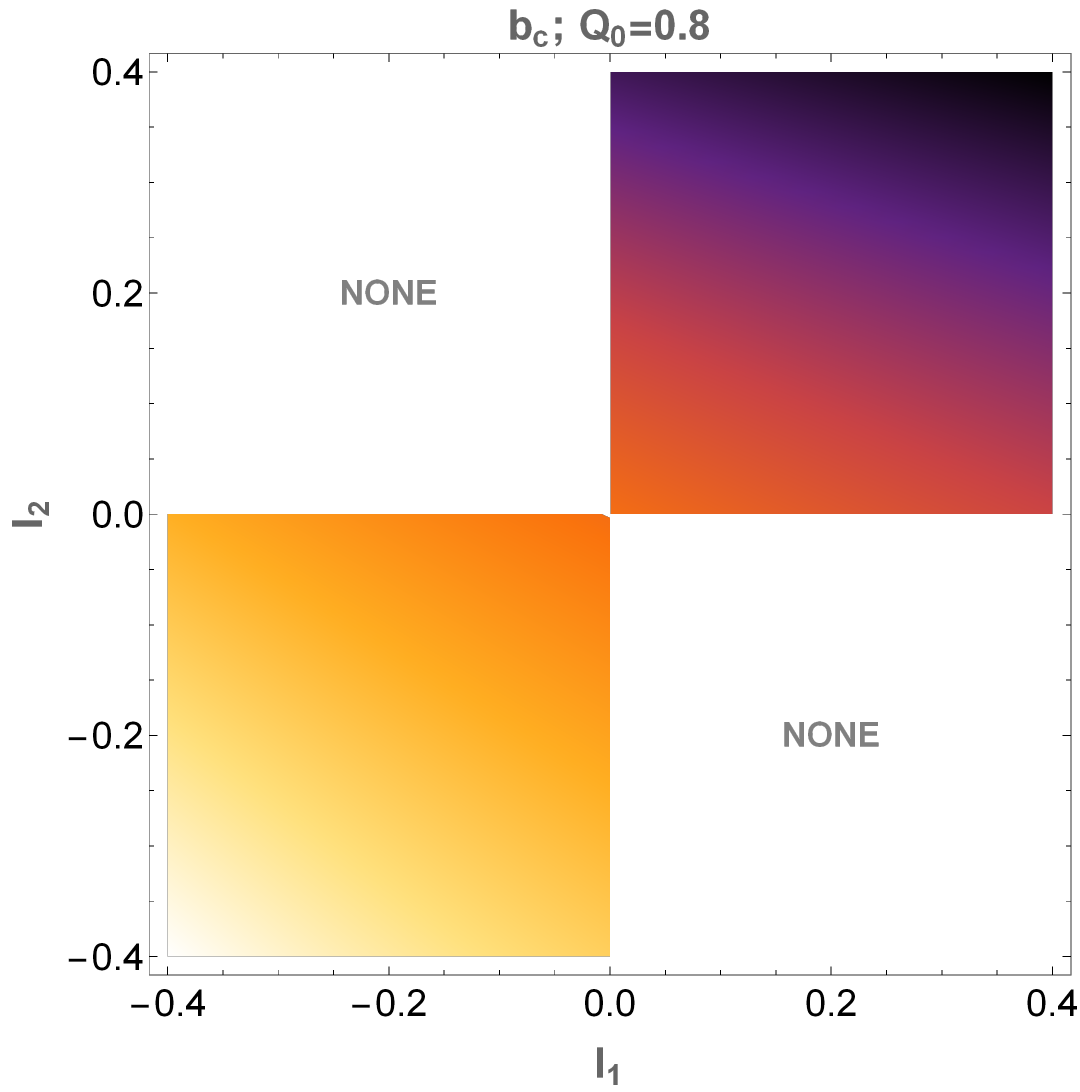}
    \end{minipage}%
    \begin{minipage}{0.09\linewidth}
        \includegraphics[width=\linewidth]{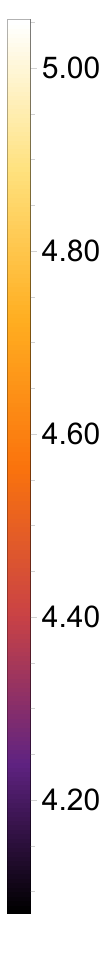}
    \end{minipage}
\end{minipage}%
\begin{minipage}{0.33\textwidth}
    \centering
    \begin{minipage}{0.9\linewidth}
        \includegraphics[width=\linewidth]{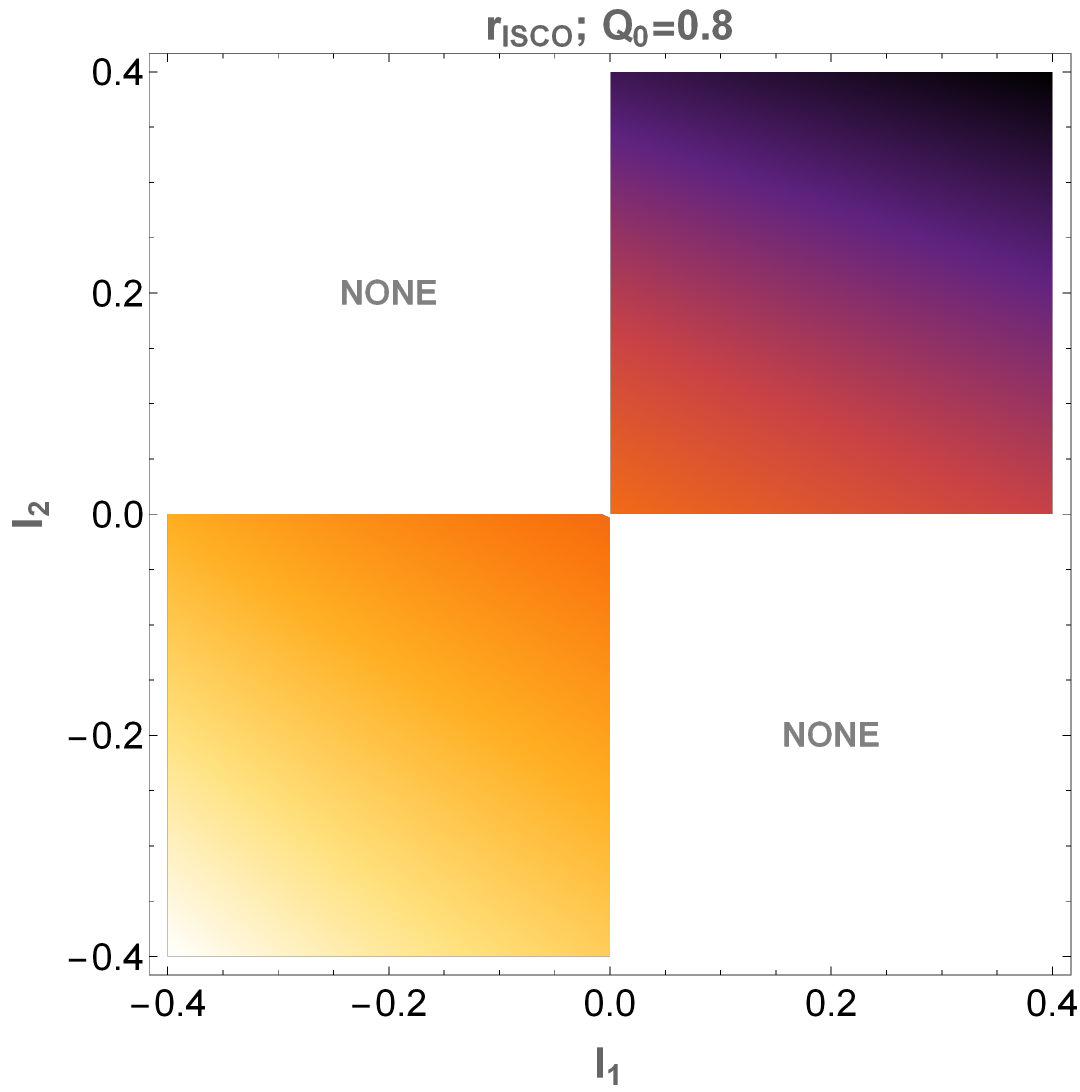}
    \end{minipage}%
    \begin{minipage}{0.09\linewidth}
        \includegraphics[width=\linewidth]{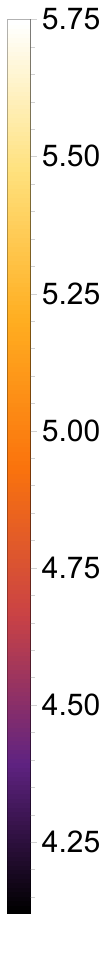}
    \end{minipage}
\end{minipage}

\caption{Dependence of the characteristic orbital quantities on the Lorentz-violating parameters $\l_1$ and $\l_2$, with $\l_1\l_2\geq0$. From left to right, the panels show the photon circular orbit radius $r_{\rm ph}$, the critical impact parameter $b_c$, and the ISCO radius $r_{\rm ISCO}$. The upper and lower rows correspond to $Q_0=0.4$ and $Q_0=0.8$, respectively, with $M=1$.}
\label{fig:2}
\end{figure*}

Using Eqs.~(\ref{eq20}), (\ref{eq21}), and (\ref{eq27}), we evaluate the photon circular orbit radius $r_{\rm ph}$, the critical impact parameter $b_c$, and the ISCO radius $r_{\rm ISCO}$ for different combinations of the Lorentz-violating parameters. Fig. ~\ref{fig:2} shows these quantities for two charge parameters, $Q_0=0.4$ and $Q_0=0.8$, with $M=1$. Within the considered parameter region, all three characteristic quantities decrease as either $\l_1$ or $\l_2$ increases. Increasing $Q_0$ produces a similar inward shift. This behavior originates from the charge-dependent term$\frac{2(1+\l_1+\l_2)Q_0^2}{2+\l_1}$
in the metric function $A(r)$, through which the charge parameter and the Lorentz-violating parameters jointly modify the null and timelike orbital structures.

\section{Characteristic orbital frequencies}
\label{sec:4}

In the preceding sections, we investigated the influence of the model parameters on the characteristic orbital structure of the charged bumblebee black hole in geometrized units. We now turn to the characteristic frequencies associated with particle motion in the vicinity of the black hole, which provide a direct link between the strong-field dynamics and potential observational signatures. Owing to the spherical symmetry of the spacetime, the vertical epicyclic frequency coincides with the azimuthal orbital frequency,
$\Omega_\theta=\Omega_\phi$,
and therefore the nodal-precession frequency vanishes
~\cite{Kluzniak:2013,Rayimbaev:2021kjs,Stuchlik:2022xtq}.
Consequently, the azimuthal and radial epicyclic frequencies constitute the two independent dynamical frequencies relevant for the orbital motion. In the following, we derive these characteristic frequencies and construct the periastron-precession frequency within the relativistic precession model, which will be used to analyze the QPO properties of X-ray binaries.

For a massive particle moving along an equatorial circular orbit, the Keplerian, or equivalently azimuthal, angular frequency measured with respect to the coordinate time $t$ is defined as $\Omega_{\phi}=\frac{d\phi}{dt}.$
Using the conserved quantities defined in Eq.~(\ref{eq17})
$E=A(r)\dot{t}$ and $L=r^2\dot{\phi}$, the angular frequency can be written as
\begin{equation}
\Omega_{\phi}
=
\frac{\dot{\phi}}{\dot{t}}
=
\frac{A(r)L}{E r^2}.
\label{eq28}
\end{equation}

Substituting the energy and angular momentum of circular orbits given in Eq.~(\ref{eq24}), the azimuthal orbital angular frequency is obtained as

\begin{equation}
\Omega_{\phi}
=
\sqrt{\frac{A'(r)}{2r}}
=
\sqrt{
\frac{M}{r^{3}}
-
\frac{2(1+\l_1+\l_2)Q_0^{2}}
{(2+\l_1)r^{4}}
}.
\label{eq29}
\end{equation}

The radial epicyclic frequency characterizes small radial oscillations around a stable circular orbit. We introduce a small radial perturbation as
\begin{equation}
r=r_c+\delta r,
\qquad
|\delta r|\ll r_c ,
\label{eq30}
\end{equation}
where $r_c$ denotes the radius of the unperturbed circular orbit. Linearizing the radial equation in Eq.~(\ref{eq22}) around the circular orbit $r=r_c$ and expanding the effective potential, where $V_{\rm eff}'(r_c)=0$, yields, to linear order in $\delta r$,
\begin{equation}
\frac{d^2\delta r}{d\tau^2}
+
\frac{V_{\rm eff}''(r_c)}
{2(1+\l_1+\l_2)}
\delta r
=0 .
\label{eq31}
\end{equation}
The curvature of the effective potential therefore determines the radial stability of the circular orbit. Using $dt/d\tau=E/A(r_c)$ and the circular-orbit energy given in Eq.~(\ref{eq24}), the radial epicyclic angular frequency measured with respect to the coordinate time is obtained as

\begin{equation}
\begin{aligned}
\Omega_r
&=
\sqrt{
\frac{2A(r)-rA'(r)}
{4(1+\l_1+\l_2)}
\left.
\frac{d^2}{dr^2}
\left[
A(r)
\left(
1+\frac{L^2}{r^2}
\right)
\right]
\right|_{r=r_c}
}
\\
&=
\Bigg\{
\frac{1}{1+\l_1+\l_2}
\Bigg[
\frac{M}{r^3}
-\frac{6M^2}{r^4}
+\frac{18M(1+\l_1+\l_2)Q_0^2}
{(2+\l_1)r^5}
\\
&\hspace{3.2cm}
-\frac{16(1+\l_1+\l_2)^2Q_0^4}
{(2+\l_1)^2r^6}
\Bigg]
\Bigg\}^{1/2}.
\label{eq32}
\end{aligned}
\end{equation}

The characteristic frequencies obtained above are derived under the geometrized-unit convention $G=c=1$. To compare these theoretical predictions with observations, we next employ the relativistic precession model and restore the corresponding physical units.
In the relativistic precession model, the observed QPO frequencies are associated with different modes of particle motion in the strong gravitational field. For a spherically symmetric spacetime, the upper-frequency peak is identified with the azimuthal orbital frequency, whereas the lower-frequency peak corresponds to the periastron-precession frequency produced by the difference between the azimuthal and radial epicyclic motions. Therefore,
\begin{equation}
\nu_{\rm U}
=
\nu_{\phi},
\label{eq33}
\end{equation}
and
\begin{equation}
\nu_{\rm L}
=
\nu_{\rm per}
=
\nu_{\phi}-\nu_r .
\label{eq34}
\end{equation}

Since the above expressions were obtained under the geometrized-unit convention, the conversion to physical frequencies requires restoring the dimensional factor associated with the black-hole mass. For a black hole with physical mass $M_{\rm BH}$, the frequency measured by a distant observer is given by
\begin{equation}
\nu_i
=
\frac{c^3}{2\pi G M_{\rm BH}}
\Omega_i ,
\qquad
i=\phi,r ,
\label{eq35}
\end{equation}
where $\Omega_i$ denotes the corresponding dimensionless angular frequency derived in the previous analysis. Here $c=2.998\times10^{8}\ {\rm m\,s^{-1}}$ and
$G=6.674\times10^{-11}\ {\rm m^{3}\,kg^{-1}\,s^{-2}}$.
The physical black-hole mass is written as
$M_{\rm BH}=m\,M_{\odot}$, where $m$ denotes the black-hole mass in solar-mass units and
$M_{\odot}=1.989\times10^{30}\ {\rm kg}$.
With these conversions, the dimensionless angular frequencies obtained from Eqs.~(\ref{eq29}) and (\ref{eq32}) are converted into physical frequencies measured in Hz.

\begin{figure*}[t!]
    \centering
    
    \begin{minipage}[b]{0.32\textwidth}
        \centering
        \includegraphics[width=\textwidth]{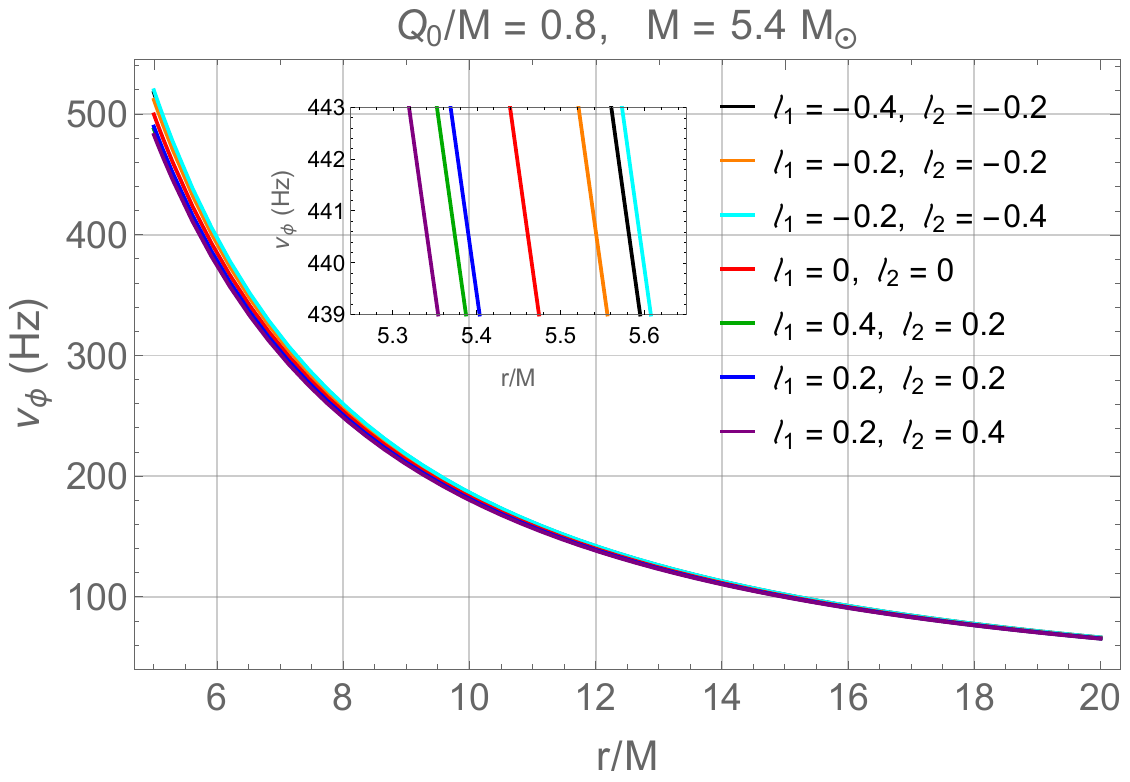}
    \end{minipage}%
    \begin{minipage}[b]{0.32\textwidth}
        \centering
        \includegraphics[width=\textwidth]{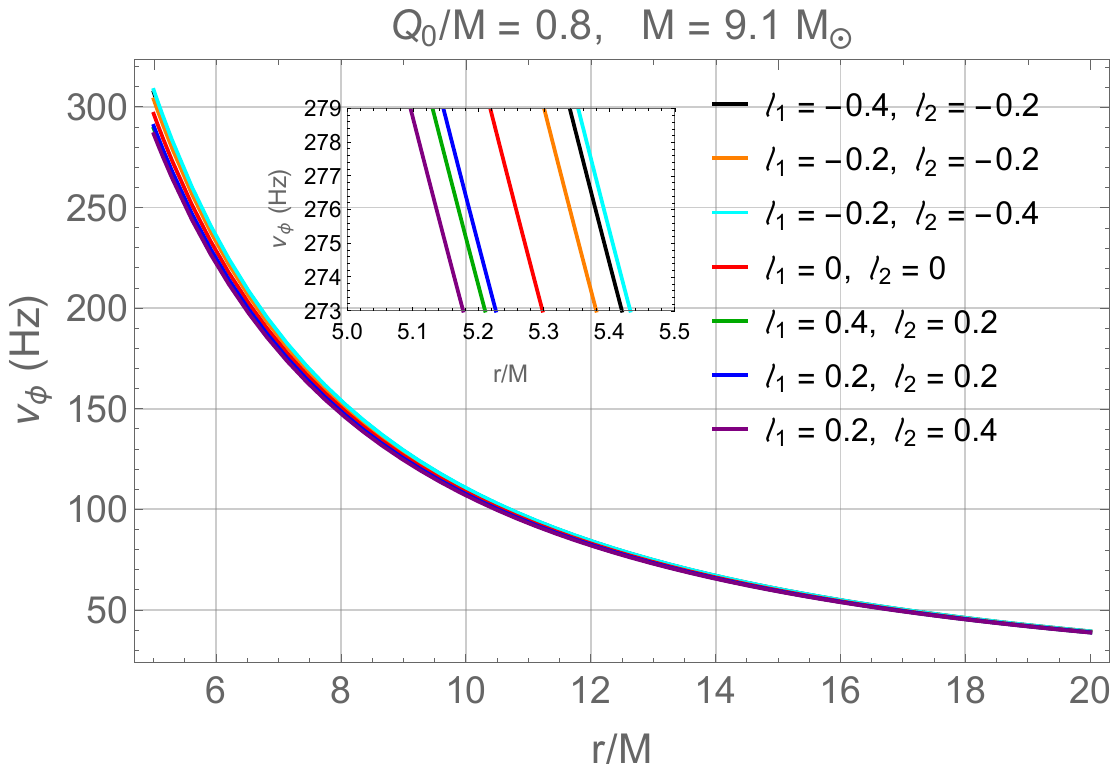}
    \end{minipage}%
    \begin{minipage}[b]{0.32\textwidth}
        \centering
        \includegraphics[width=\textwidth]{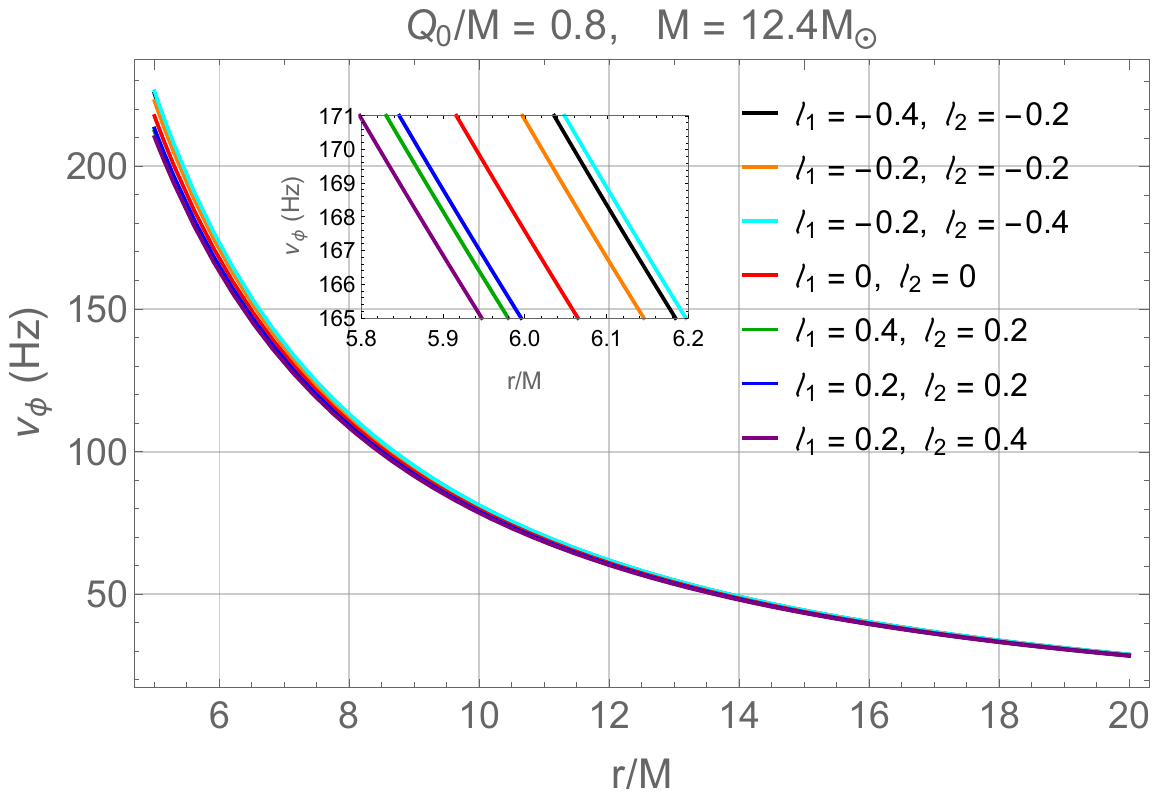}
    \end{minipage}

    \vspace{0.3cm}

    \begin{minipage}[b]{0.32\textwidth}
        \centering
        \includegraphics[width=\textwidth]{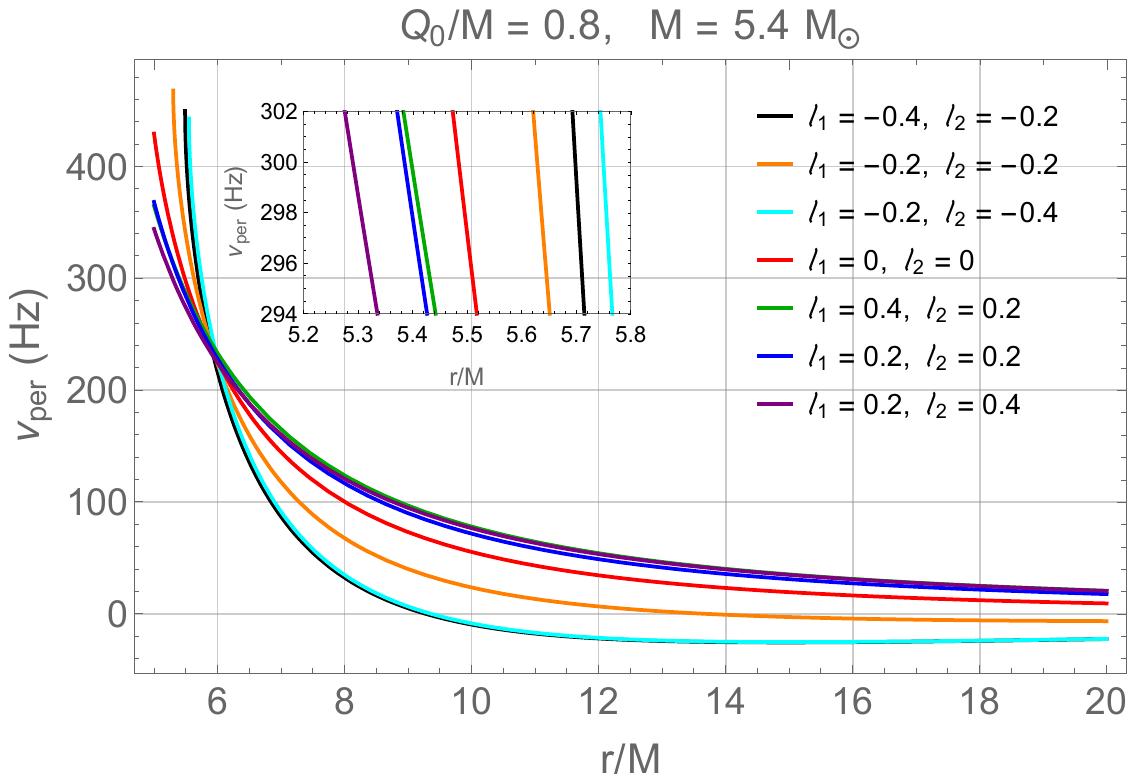}
    \end{minipage}%
    \begin{minipage}[b]{0.32\textwidth}
        \centering
        \includegraphics[width=\textwidth]{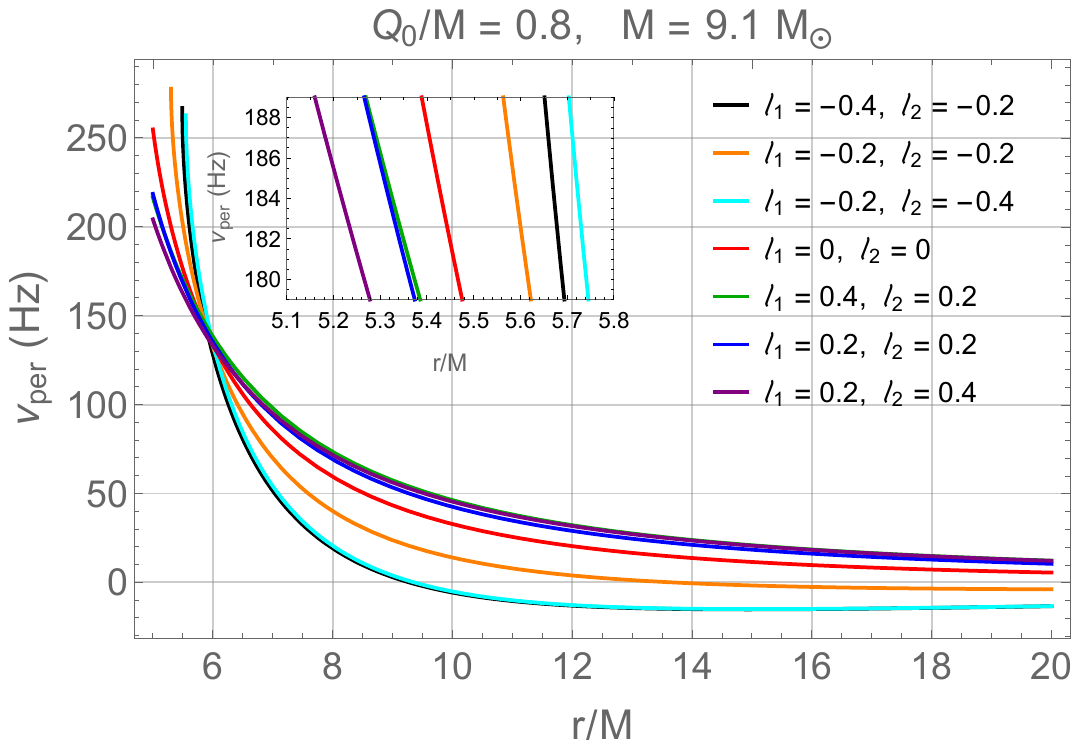}
    \end{minipage}%
    \begin{minipage}[b]{0.32\textwidth}
        \centering
        \includegraphics[width=\textwidth]{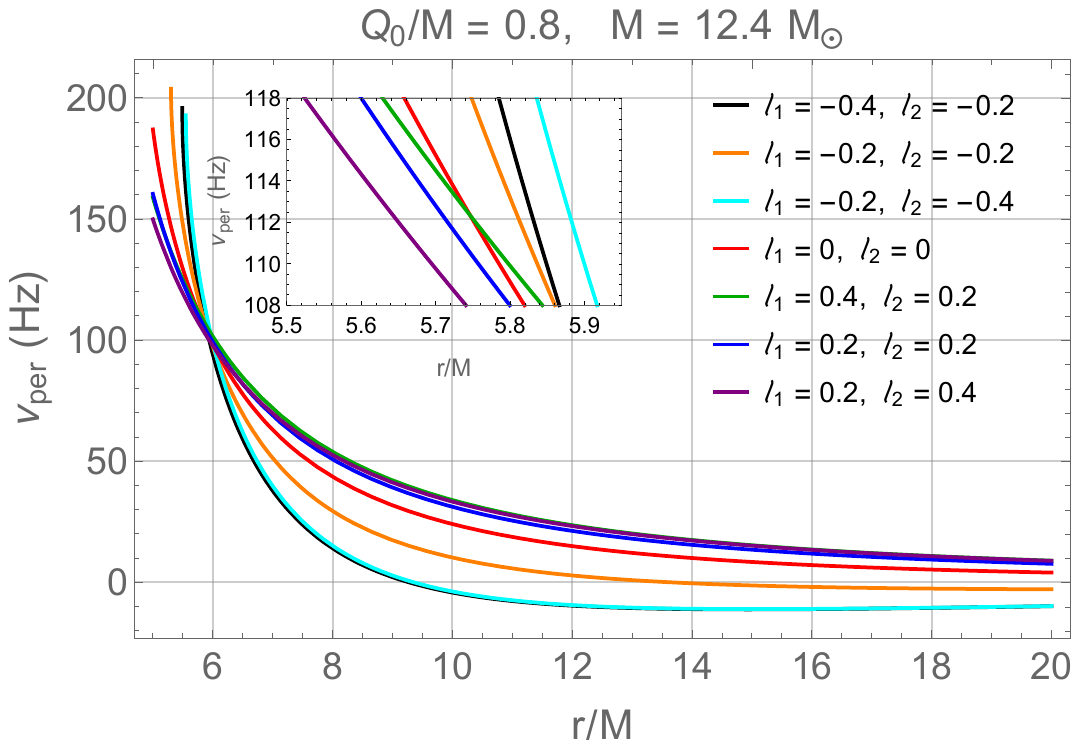}
    \end{minipage}

     \vspace{0.3cm}

    \begin{minipage}[b]{0.32\textwidth}
        \centering
        \includegraphics[width=\textwidth]{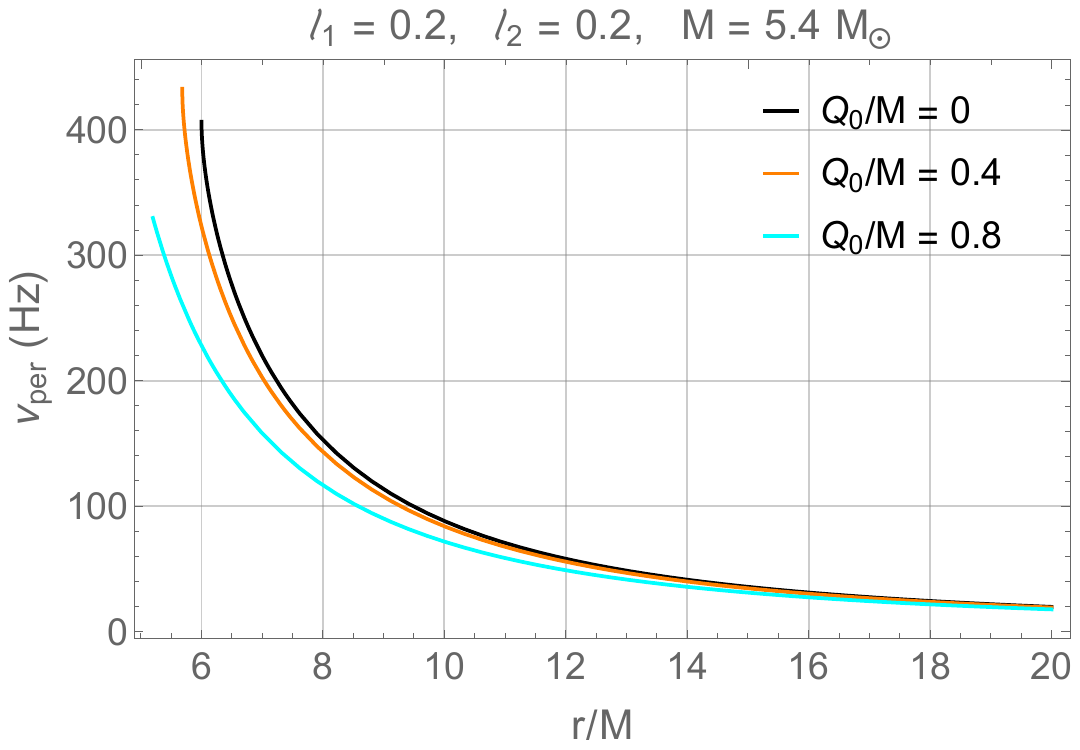}
    \end{minipage}%
    \begin{minipage}[b]{0.32\textwidth}
        \centering
        \includegraphics[width=\textwidth]{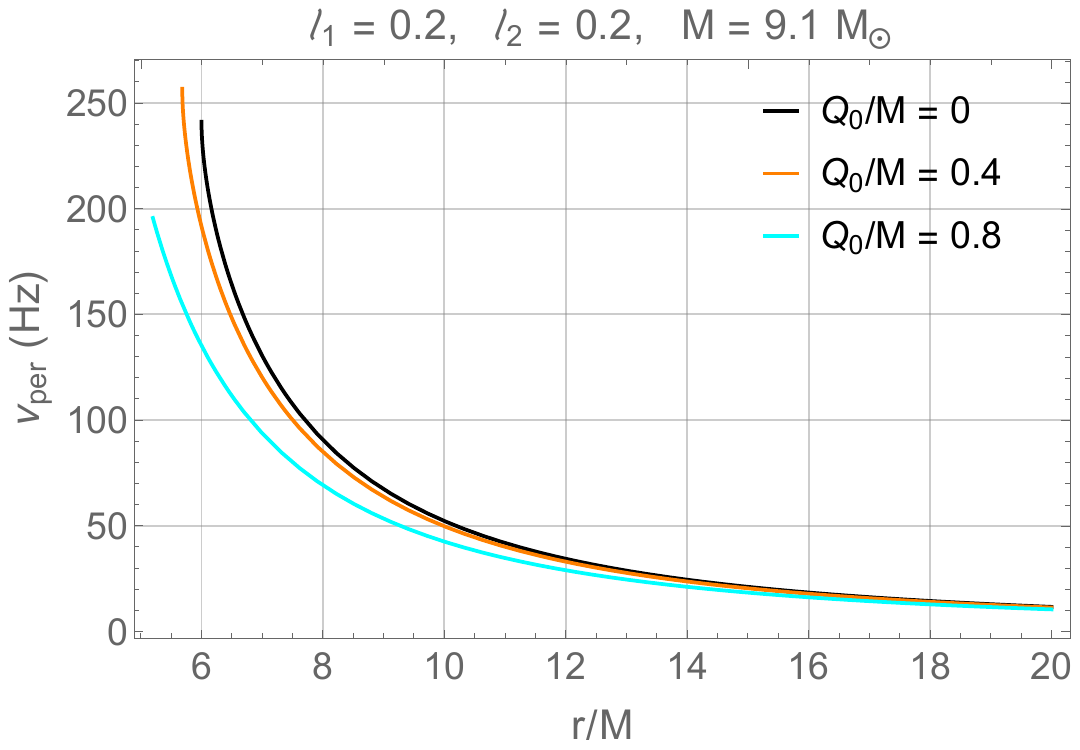}
    \end{minipage}%
    \begin{minipage}[b]{0.32\textwidth}
        \centering
        \includegraphics[width=\textwidth]{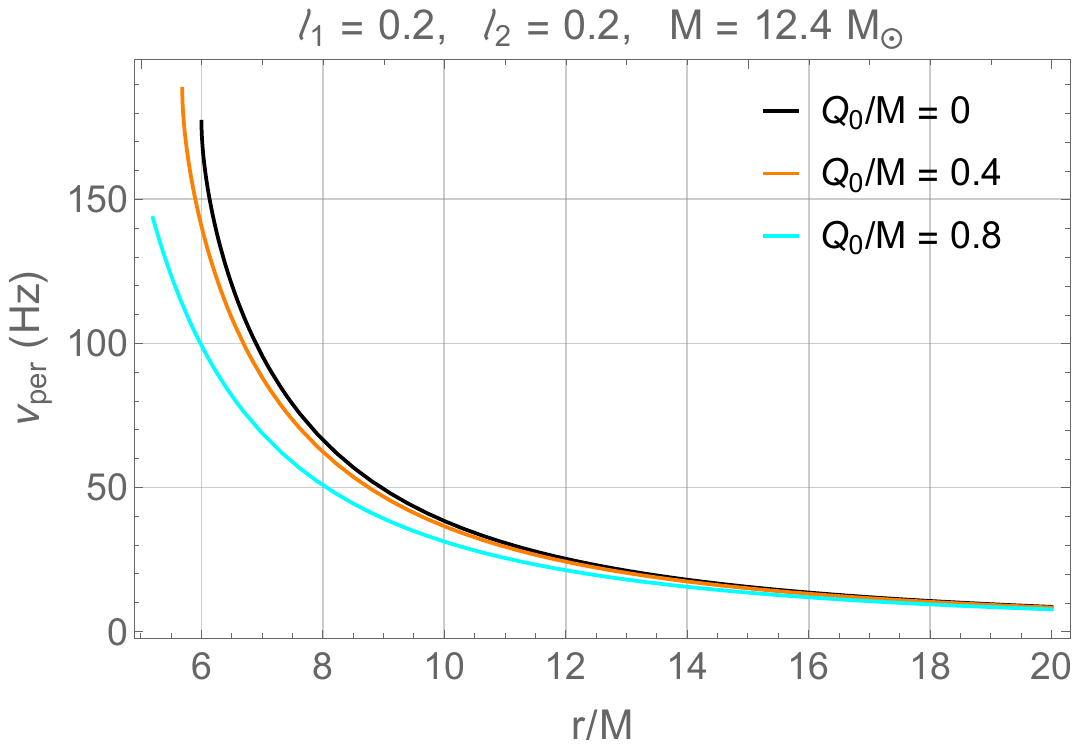}
    \end{minipage}

\caption{Azimuthal orbital frequency $\nu_{\phi}$ and periastron-precession frequency $\nu_{\rm per}$ versus dimensionless orbital radius $r/M$ for GRO J1655--40, XTE J1550--564, and GRS 1915+105 (columns left to right). The black hole masses $M$ are fixed to their best-fit values in Table~\ref{tab1}. The top and middle panels display $\nu_{\phi}$ and $\nu_{\rm per}$ for $Q_0/M=0.8$ with varying $(\l_1,\l_2)$. The bottom panel shows $\nu_{\rm per}$ for fixed $\l_1=\l_2=0.2$ and varying $Q_0/M$. Insets zoom in on the observed frequency ranges in Table~\ref{tab1}.}
\label{fig:3}

\end{figure*}

We calculated the characteristic frequencies of charged bumblebee black holes in physical units and studied the effects of Lorentz violation parameters and charge parameters on the distribution of orbital frequencies. By examining different combinations of $\l_1$, $\l_2$, and $Q_0$, we demonstrated their influence on the azimuthal orbital frequency,  and periastron-precession frequency
The range of Lorentz violation parameters is $-0.4 \leq \l_1, \l_2 \leq 0.4$.
It should be emphasized that this range does not comply with the current observational limits on Lorentz violation. The existing weak-field tests and astrophysical observations typically require the Lorentz violation parameters to be much smaller ~\cite{Casana:2017jkc,QiQi:2026pnb}. Therefore, the parameter range adopted here is used as a phenomenological window to amplify possible strong-field effects and study how Lorentz violation alters the characteristic frequencies around the black hole. This choice enables us to explore the qualitative behavior of the model in the strong-field region.

Fig.~\ref{fig:3} shows the frequency--radius relations for GRO J1655--40, XTE J1550--564, and GRS 1915+105 from left to right, with the black hole masses fixed at the central values listed in Table~\ref{tab1}. In the top and middle rows, we fix $Q_0/M=0.8$ and vary the same-sign combinations of $(\l_1,\l_2)$ indicated in the legends. The azimuthal orbital frequency $\nu_{\phi}$ decreases as $r/M$ increases, while the periastron-precession frequency $\nu_{\rm per}$ generally decreases over the radial regions relevant to the observed QPOs. At a fixed radius, increasing $\l_1$ or $\l_2$ while holding the other parameters fixed leads to a slightly lower $\nu_{\phi}$ for the parameter values shown. The $\nu_{\phi}$ curves therefore remain closely spaced over most of the plotted radial range, whereas the $\nu_{\rm per}$ curves exhibit more pronounced differences. This indicates that, for the illustrated configurations, $\nu_{\rm per}$ is more sensitive to changes in the Lorentz-violating parameters than $\nu_{\phi}$.The $\nu_{\rm per}$ curves intersect: curves with larger Lorentz‑violation parameters sit lower before crossing and flip their relative positions afterwards.

In the bottom row, we fix $\l_1=\l_2=0.2$ and vary $Q_0/M$. For the illustrated choices, increasing $Q_0/M$ lowers $\nu_{\rm per}$ at a given $r/M$; the orbital‑frequency expression likewise predicts a lower $\nu_{\phi}$ when $Q_0/M$ increases while the other parameters are held fixed (Eq.~\ref{eq29}).
The insets enlarge the regions near the observed upper and lower QPO frequencies listed in Table~\ref{tab1}. At $Q_0/M=0.8$, the theoretical curves reach frequencies near the observed values for all three sources at radii of approximately $r/M\sim5$--$6$.These qualitative results motivate a  comparison with the observed upper and lower QPO frequencies of each source at a common orbital radius. 

\section{MCMC constraints from X-ray binary QPO observations}
\label{sec:5}

In the previous section, we derived the characteristic orbital frequencies of the charged bumblebee black hole and constructed the QPO frequencies within the relativistic precession model. In this section, we use the observed QPO frequencies from X-ray binary systems to constrain the effective parameter space of the model.
By comparing the theoretical frequencies with observations, we perform a Bayesian parameter estimation using MCMC method implemented with the Python package \texttt{emcee}~\cite{Foreman-Mackey:2012any,Ali:2026pmh}. The posterior distributions and allowed ranges of the effective parameters entering the QPO model are then obtained.

The QPO observations used in this analysis are taken from three representative stellar-mass black hole X-ray binary systems, including GRO J1655--40~\cite{Motta:2013wga}, XTE J1550--564~\cite{Orosz:2011ki,Remillard:2002cy}, and GRS 1915+105~\cite{Reid:2014ywa,Remillard:2006fc}. These systems exhibit well-measured twin-peak QPOs and provide suitable observational samples for constraining the effective parameters of the charged bumblebee black hole. The corresponding black hole masses and observed QPO frequencies adopted in our analysis are summarized in Table~\ref{tab1}. 

For a static spherically symmetric spacetime, the relativistic precession model describes the observed twin-peak QPOs through two independent dynamical frequencies, namely the azimuthal orbital frequency $\nu_{\phi}$ and the periastron-precession frequency $\nu_{\rm per} $. Therefore, for an individual X-ray binary system, the available QPO observations provide a limited number of independent frequency constraints. In the charged bumblebee black hole considered in this work, the original parameter space contains five quantities

\begin{equation}
\left\{
M,\frac{r_{\rm }}{M},\l_1,\l_2,\frac{Q_0}{M}
\right\}
\label{eq36}
\end{equation}

where $M$ is the black hole mass, $r_{\rm }/M$ denotes the dimensionless radius of the QPO-emitting region, while $\l_1$, $\l_2$, and $Q_0/M$ describe the Lorentz-violating and charge-related corrections.

\begin{table}[h]
\centering
\caption{The observational parameters of black hole X-ray binaries used for QPO constraints.}
\begin{tabular}{c c c c}
\toprule
 & GRO J1655--40 & XTE J1550--564 & GRS 1915+105 \\
\midrule

$M(M_\odot)$ 
& $5.4\pm0.3$ ~\cite{Motta:2013wga}
& $9.1\pm0.61$ ~\cite{Orosz:2011ki,Remillard:2002cy}
& $12.4^{+2.0}_{-1.8}$ ~\cite{Reid:2014ywa,Remillard:2006fc}
\\

$\nu_{\phi}({\rm Hz})$
& $441\pm2$ ~\cite{Motta:2013wga}
& $276\pm3$ ~\cite{Orosz:2011ki,Remillard:2002cy}
& $168\pm3$ ~\cite{Reid:2014ywa,Remillard:2006fc}
\\

$\nu_{\rm per}({\rm Hz})$
& $298\pm4$ ~\cite{Motta:2013wga}
& $184\pm5$ ~\cite{Orosz:2011ki,Remillard:2002cy}
& $113\pm5$ ~\cite{Reid:2014ywa,Remillard:2006fc}
\\

\bottomrule
\end{tabular}
\label{tab1}
\end{table}

As shown by the frequency expressions derived in Sec.~\ref{sec:4}, $\l_1$, $\l_2$, and $Q_0/M$ enter the predicted QPO frequencies through particular combinations. Consequently, different values of the original parameters can produce the same pair of frequencies. 
Therefore, directly sampling the original parameter set would introduce significant degeneracies. To extract the physically relevant combinations constrained by QPO observations, we introduce the following effective parameters

\begin{equation}
C \equiv 1+\l_1+\l_2,\qquad
\beta \equiv \frac{2(1+\l_1+\l_2)}{2+\l_1}\left(\frac{Q_0}{M}\right)^2.
\label{eq37}
\end{equation}

The predicted QPO frequencies can be parameterized by the four quantities $(M,X,C,\beta)$ instead of the original five, because $C$ and $\beta$ represent combinations of the original parameters rather than additional free parameters. Here, $C$ enters the radial epicyclic frequency through the radial metric factor, while $\beta$ acts as an effective charge coefficient analogous to that in Reissner--Nordstr\"om geometry. This reparameterization removes one redundant direction from the frequency model, but does not add an observational constraint: two measured QPO frequencies cannot, on their own, uniquely determine all four effective parameters. Accordingly, marginalized constraints on $M$, $X$, $C$, and $\beta$ depend on the adopted priors and physical bounds, and any constraints on $C$ and $\beta$ cannot be interpreted as separate measurements of $\l_1$, $\l_2$, and $Q_0/M$.

Meanwhile, the dimensionless radial coordinate of the QPO-emitting region is defined as:$ X=\frac{r}{M}$
With these definitions, the metric function can be rewritten in the dimensionless form

\begin{equation}
A(X)=1-\frac{2}{X}+\frac{\beta}{X^2}
\label{eq38}
\end{equation}

Accordingly, the azimuthal orbital frequency obtained from Eq.~(\ref{eq29}) becomes

\begin{equation}
(M\Omega_{\phi})^2
=
\frac{X-\beta}{X^4}
\label{eq39}
\end{equation}

while the radial epicyclic frequency in Eq.~(\ref{eq32}) can be expressed as

\begin{equation}
(M\Omega_r)^2=\frac{X^3-6X^2+9\beta X-4\beta^2}{CX^6}
\label{eq40}
\end{equation}

It is therefore clear that the QPO frequencies depend on the effective parameter set
\begin{equation}
\Theta=(M,X,C,\beta)
\label{eq41}
\end{equation}
The subsequent MCMC analysis is performed in the effective parameter space $\Theta=(M,X,C,\beta)$, which avoids the degeneracy among the original parameters and directly constrains the physical combinations governing the orbital dynamics and QPO frequencies.

\begin{table*}[t]
\centering
\caption{The priors for the effective parameters. The entries for $M$, $X$, and $\beta$ give the centre and lower/upper scale parameters of split Gaussian priors; $C$ has a uniform prior.}
\begin{tabular}{lcccc}
\hline
Source & Prior $M/M_{\odot}$ & Prior $X=r/M$ & Prior $C$ & Prior $\beta$ \\
\hline
GRO J1655--40
& $5.1596^{+0.2992}_{-0.2976}$
& $5.4793^{+0.0255}_{-0.0282}$
& $\mathcal{U}(0.2,1.8)$
& $0.6610^{+0.0141}_{-0.0118}$ \\

XTE J1550--564
& $9.255^{+0.5891}_{-0.5445}$
& $5.2641^{+0.02633}_{-0.03204}$
& $\mathcal{U}(0.2,1.8)$
& $0.7736^{+0.0151}_{-0.0113}$ \\

GRS 1915+105
& $12.1825^{+0.0298}_{-0.0299}$
& $5.9431^{+0.01555}_{-0.01613}$
& $\mathcal{U}(0.2,1.8)$
& $0.4497^{+0.0135}_{-0.0129}$ \\
\hline
\end{tabular}
\label{tab2}
\end{table*}

\begin{table*}[t]
\centering
\caption{Posterior medians and 68\% central credible intervals for the effective parameters.}
\begin{tabular}{lcccc}
\hline
Source & Posterior $M/M_{\odot}$ & Posterior $X=r/M$ & Posterior $C$ & Posterior $\beta$ \\
\hline
GRO J1655--40
& $5.3534^{+0.0449}_{-0.0432}$
& $5.4807^{+0.0258}_{-0.0269}$
& $1.0136^{+0.0845}_{-0.0780}$
& $0.6632^{+0.0133}_{-0.0125}$ \\

XTE J1550--564
& $8.9800^{+0.1199}_{-0.1170}$
& $5.2597^{+0.0279}_{-0.0297}$
& $1.0342^{+0.1497}_{-0.1255}$
& $0.7764^{+0.0138}_{-0.0125}$ \\

GRS 1915+105
& $12.1935^{+0.0292}_{-0.0294}$
& $5.9522^{+0.0153}_{-0.0152}$
& $0.9184^{+0.1794}_{-0.1395}$
& $0.4528^{+0.0133}_{-0.0130}$ \\
\hline
\end{tabular}
\label{tab:posterior}
\end{table*}

For a given set of parameters $\Theta$, the theoretical upper and lower QPO frequencies are calculated from the azimuthal orbital frequency and the periastron-precession frequency derived in Sec.~\ref{sec:4}, namely $\nu_{\rm U}^{\rm th}=\nu_{\phi},$ and $\nu_{\rm L}^{\rm th}=\nu_{\phi}-\nu_r.$ 
The posterior probability distribution is obtained according to Bayes' theorem~\cite{Liu:2023vfh,Sharma:2017wfu},

\begin{equation}
P(\Theta|D,\mathcal{M})
=
\frac{
\mathcal{L}(D|\Theta,\mathcal{M})
\pi(\Theta|\mathcal{M})
}
{P(D|\mathcal{M})}
\label{eq42}
\end{equation}

where $\mathcal{L}(D|\Theta,\mathcal{M})$ and $\pi(\Theta|\mathcal{M})$ denote the likelihood function and prior distribution under the model $\mathcal{M}$, respectively.The likelihood function is constructed by comparing the theoretical QPO frequencies with the observed values. For a given X-ray binary system, it is defined as

\begin{equation}
\ln\mathcal{L}=-\frac{1}{2}\left[\frac{(\nu_{\rm U}^{\rm th}-\nu_{\rm U}^{\rm obs})^2}{\sigma_{\rm U}^{2}}+\frac{(\nu_{\rm L}^{\rm th}-\nu_{\rm L}^{\rm obs})^2}{\sigma_{\rm L}^{2}}\right]
\label{eq43}
\end{equation}

where $\nu_{\rm U}^{\rm obs}$ and $\nu_{\rm L}^{\rm obs}$ are the observed upper and lower QPO frequencies, while $\sigma_{\rm U}$ and $\sigma_{\rm L}$ denote the corresponding uncertainties.

The priors for $M$, $X$, and $\beta$ are informed by a preliminary Reissner--Nordstr\"om (RN)‑limit reference treatment, adopting a prior setup analogous to the procedure presented in Ref.~\cite{Zhang:2025acq}. Studies of Schwarzschild‑like bumblebee black holes motivate examining a regime of small Lorentz‑violating deformations~\cite{Casana:2017jkc,QiQi:2026pnb}. Although bounds from those one‑parameter solutions cannot be directly applied to the distinct parameters $\l_1$ and $\l_2$ of the present charged solution, they suggest adopting RN geometry as a reasonable reference for $|\l_1|,|\l_2|\ll1$. In this limit, the composite parameter $C=1+\l_1+\l_2$ approaches unity, and $\beta$ reduces to its standard Reissner--Nordstr\"om value for fixed $Q_0/M$. For each source, we fit the observed twin‑peak QPO frequencies within the RN limit $\l_1=\l_2=0$, incorporating the independent dynamical mass measurement from Table~\ref{tab1}, and use the resulting outputs to define source‑dependent split‑Gaussian prior factors for $M$, $X$, and $\beta$ in the charged bumblebee model. The corresponding central values and lower‑/upper‑scale parameters are summarized in Table~\ref{tab2}. Under the working assumption that modest Lorentz‑violating corrections do not shift the preferred parameter region far from the RN reference solution, these RN‑derived results guide the prior locations for $M$, $X$, and $\beta$; We stress that this assumption only informs our prior choice and does not constitute an independent constraint supplied by the QPO data. For $\theta_i\in\{M,X,\beta\}$, each prior factor has the form
$\pi_i(\theta_i)\propto\exp\!\left[-\frac{(\theta_i-\mu_i)^2}{2\sigma_i^2(\theta_i)}\right],\label{eq:split_gaussian_prior}$
where $\sigma_i(\theta_i)=\sigma_{i,-}$ for $\theta_i<\mu_i$ and $\sigma_i(\theta_i)=\sigma_{i,+}$ for $\theta_i\geq\mu_i$.

\begin{figure}[!t]
    \centering

    \includegraphics[width=0.8\columnwidth]{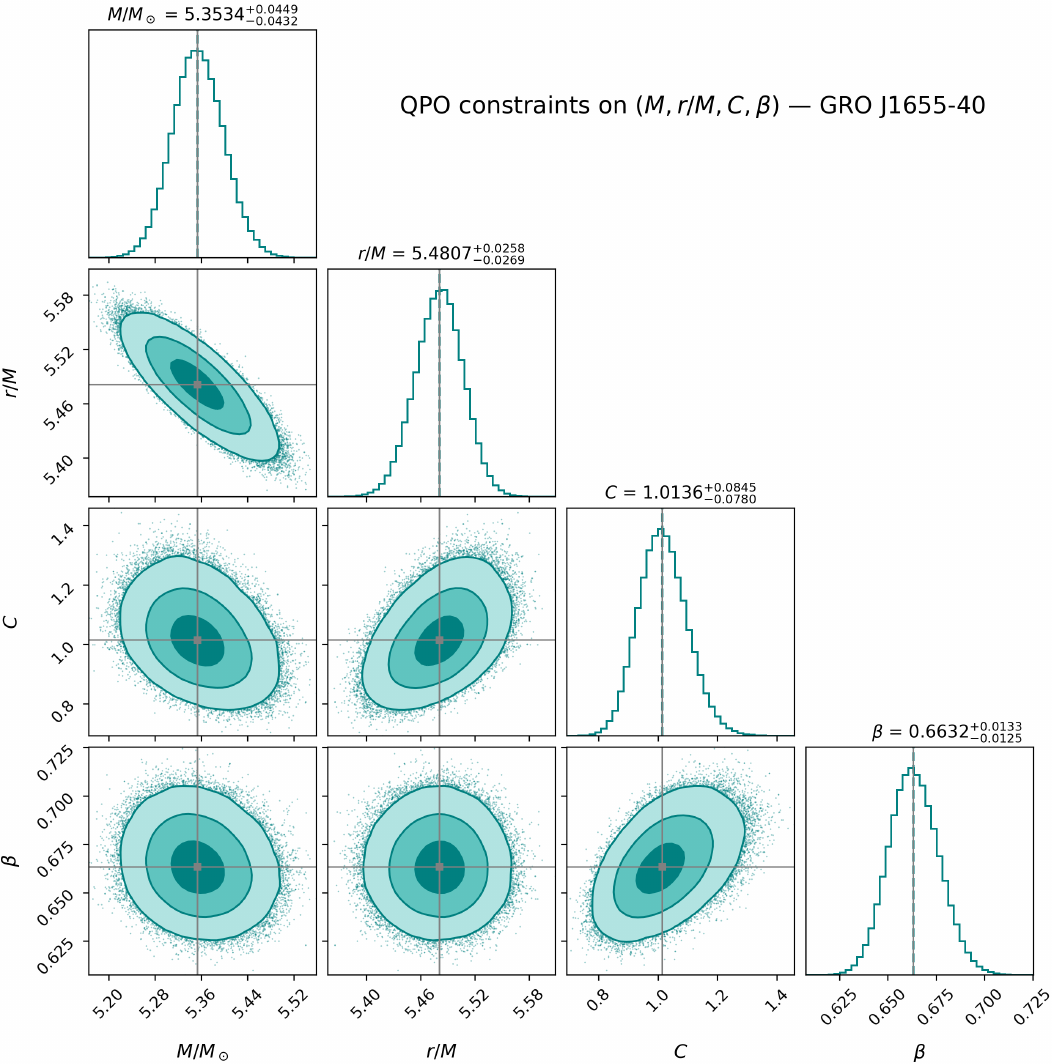}

    \vspace{-0.6em}

    \includegraphics[width=0.8\columnwidth]{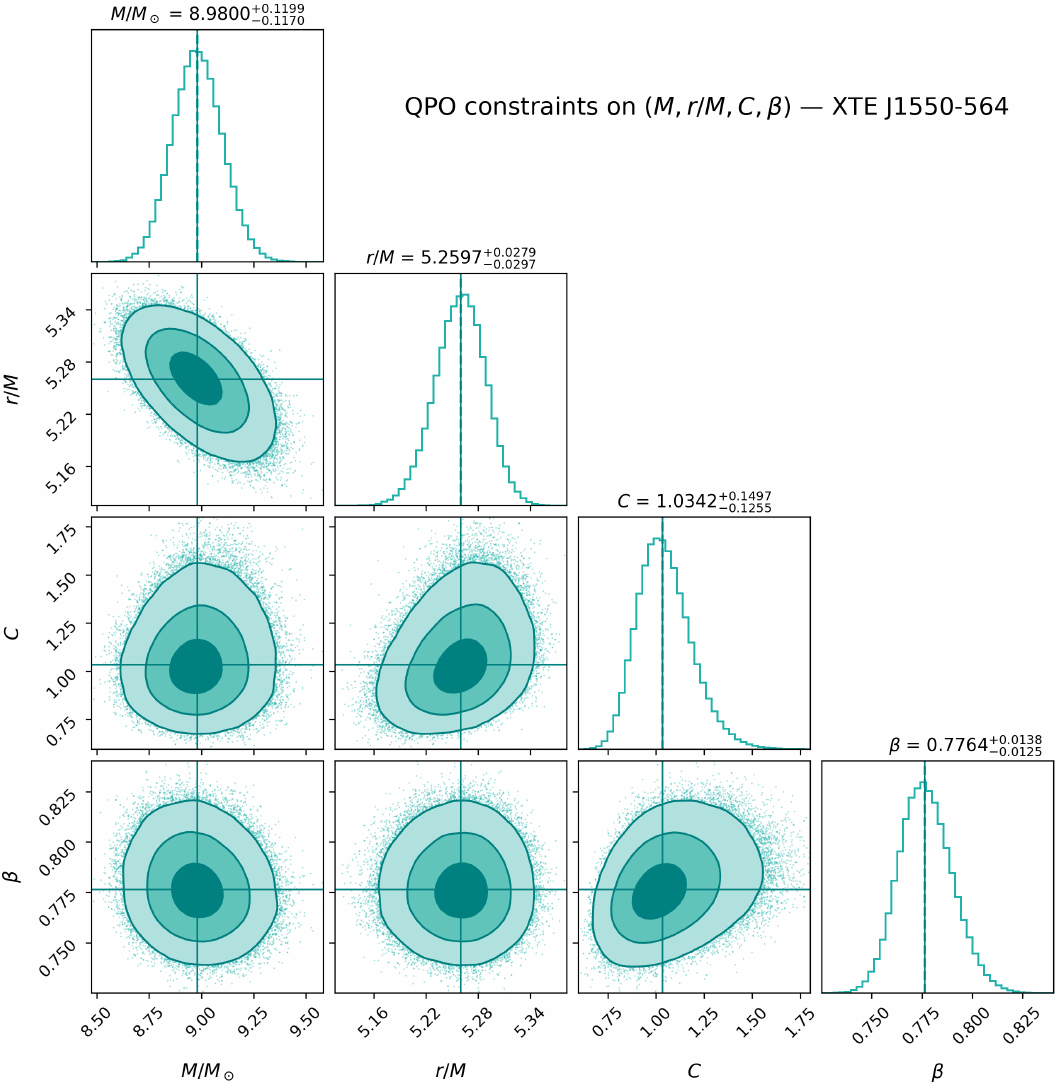}

    \vspace{-0.6em}

    \includegraphics[width=0.8\columnwidth]{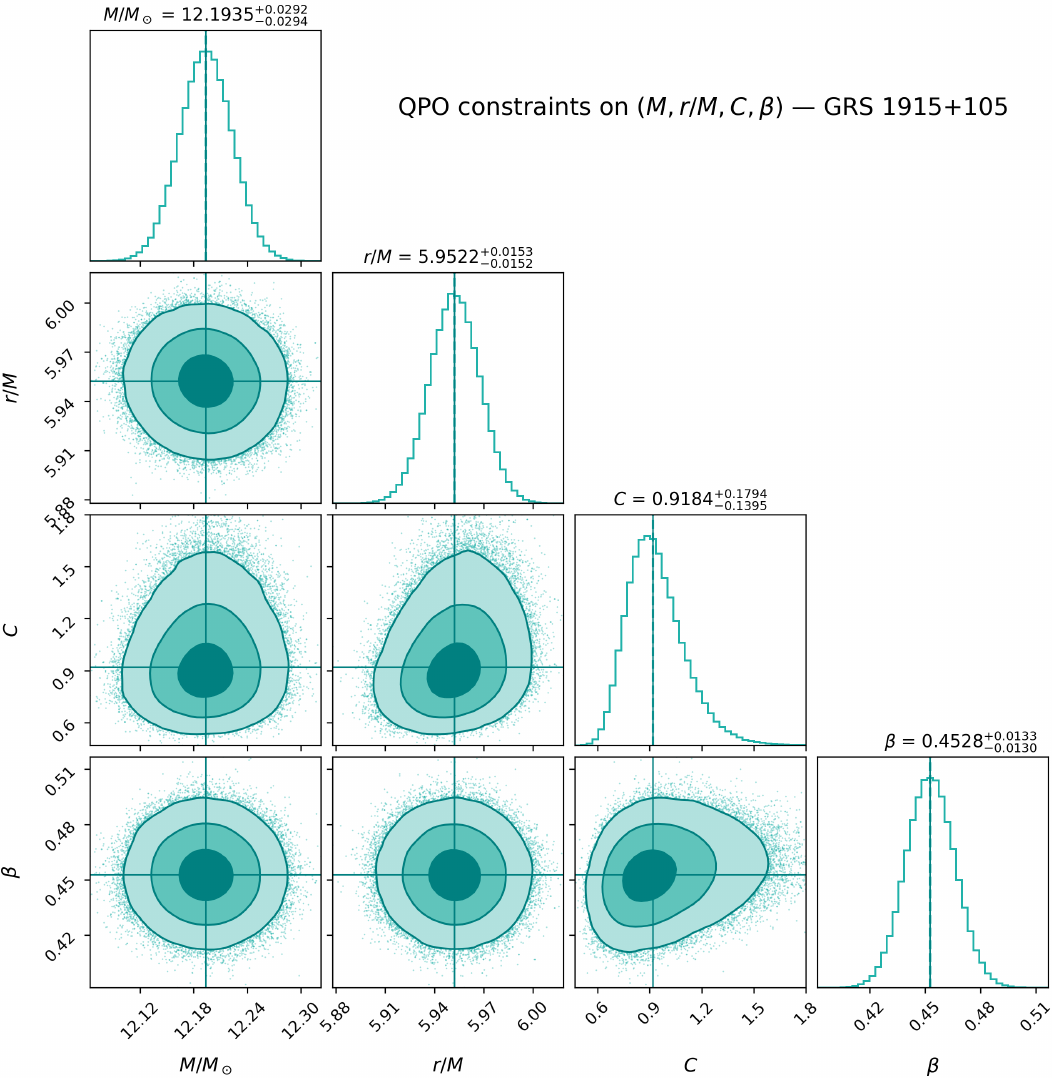}

   \caption{Posterior distributions of $\Theta=(M,X,C,\beta)$ from the RP-QPO analysis. From top to bottom: GRO J1655--40, XTE J1550--564, and GRS 1915+105.}
    \label{fig:4}
\end{figure}

The marginalized posterior distributions and parameter correlations are shown in Fig.~\ref{fig:4}. From top to bottom, the corner plots correspond to GRO J1655--40, XTE J1550--564, and GRS 1915+105. The diagonal panels display the one-dimensional marginal distributions of $M$, $X$, $C$, and $\beta$, while the off-diagonal panels display their joint distributions. A pronounced anticorrelation between $M$ and $X$ is visible for GRO J1655--40 and XTE J1550--564, whereas it is much weaker for GRS 1915+105. The plots also show a positive correlation between $X$ and $C$, as well as between $C$ and $\beta$. These correlations indicate that changes in one parameter can be partly compensated by changes in another when reproducing the observed frequency pair. Consequently, the one-dimensional intervals should be interpreted together with the joint posterior distributions.

The numerical results are given in Table~\ref{tab:posterior}, and the priors used to obtain them are listed in Table~\ref{tab2}. The posterior median radii are $X=5.4807$, $5.2597$, and $5.9522$ for GRO J1655--40, XTE J1550--564, and GRS 1915+105, respectively. The corresponding median values of $C$ are $1.0136$, $1.0342$, and $0.9184$, while those of $\beta$ are $0.6632$, $0.7764$, and $0.4528$. The marginalized distribution of $C$ is narrowest for GRO J1655--40 and broadest for GRS 1915+105. Although the median $C$ lies slightly above unity for the first two sources and below unity for GRS 1915+105, all three 68\% credible intervals include $C=1$. These results therefore do not establish a preference for either sign branch of the Lorentz-violating parameters.

Comparison with Table~\ref{tab2} is essential when assessing the origin of the posterior widths. The posterior widths of $X$ and $\beta$ are very similar to their adopted split-Gaussian prior scales for all three sources. The mass posterior becomes narrower for GRO J1655--40 and XTE J1550--564, whereas for GRS 1915+105 its width remains close to that of the mass prior. The inferred intervals, particularly those of $X$ and $\beta$, must therefore be understood as conditional on the informative RN-based priors; their precision cannot be attributed solely to the charged bumblebee QPO likelihood. The uniform prior on $C$ leads to more localized marginal posteriors, but these too are conditional on the adopted priors for the other parameters.

Finally, the constraints on $C$ and $\beta$ apply to combinations of the original parameters, rather than to $\l_1$, $\l_2$, and $Q_0/M$ separately. For an admissible choice of $\l_1$, the same effective values can be obtained from$
\l_2=C-1-\l_1,\qquad
\left(\frac{Q_0}{M}\right)^2=\frac{\beta(2+\l_1)}{2C}.
$ Thus, different physically allowed triples $(\l_1,\l_2,Q_0/M)$ may correspond to the same $(C,\beta)$ and yield identical QPO predictions. The inferred effective parameter region provides a necessary consistency requirement for an underlying parameter choice within the assumed RP model, but it is not sufficient to determine that choice uniquely or to establish the charged bumblebee interpretation. More independent observational information is needed to disentangle the Lorentz-violating and charge contributions and to test the model more decisively.

\section{Conclusions}
\label{sec:6}

In this work, we have investigated the strong-field orbital dynamics of a charged bumblebee black hole and explored its observational implications through HFQPOs from black hole X-ray binary systems. By analyzing the null and timelike geodesics, we derived the photon-sphere radius, the corresponding critical impact parameter, and the ISCO radius, and examined their dependence on the Lorentz-violating and charge parameters. Within the parameter region considered, these characteristic orbital quantities shift inward as the Lorentz-violating parameters or the charge parameter increase. We further derived the azimuthal and radial epicyclic frequencies and constructed the twin-peak QPO frequencies within the relativistic precession model.
We find that the Lorentz-violating parameters and charge parameter modify the orbital dynamics through specific combinations rather than independent contributions to the observable frequencies. This intrinsic degeneracy prevents the QPO observations from independently constraining the original microscopic parameters. Therefore, we introduced an effective parameterization that captures the combinations directly relevant to the orbital dynamics and adopted it for the subsequent observational analysis.

Using Bayesian inference with the affine-invariant Markov Chain Monte Carlo method implemented with the \texttt{emcee} package, we constrained the effective parameter space with the HFQPO observations of three stellar-mass black hole X-ray binaries, GRO J1655--40, XTE J1550--564, and GRS 1915+105. The results show that the charged bumblebee black-hole spacetime can reproduce the observed HFQPO frequencies within the constrained parameter region. The inferred black hole masses remain consistent with independent dynamical measurements, while the posterior distributions of the orbital radii and effective parameters exhibit source-dependent behavior.

Accordingly, the constraints obtained here should be interpreted as bounds on the effective geometry probed by HFQPOs, rather than as direct measurements of the individual Lorentz-violating and charge parameters. Future X-ray timing observations with improved precision and larger source samples may further tighten the effective constraints, while disentangling the underlying contributions will require additional independent measurements.

\section{Acknowledgements}
This research was  supported by the National Natural Science Foundation of China (Grant No.12265007), and the Guizhou Provincial Major Scientific and Technological Program (XKBF(2025)010).
\bibliographystyle{apsrev4-2}

\begin{thebibliography}{}

\bibitem{Kostelecky:2003fs}
V.~A.~Kostelecky,
Phys. Rev. D \textbf{69} (2004), 105009
doi:10.1103/PhysRevD.69.105009
[arXiv:hep-th/0312310 [hep-th]].

\bibitem{Colladay:1998fq}
D.~Colladay and V.~A.~Kostelecky,
Phys. Rev. D \textbf{58} (1998), 116002
doi:10.1103/PhysRevD.58.116002
[arXiv:hep-ph/9809521 [hep-ph]].


\bibitem{Colladay:1996iz}
D.~Colladay and V.~A.~Kostelecky,
Phys. Rev. D \textbf{55} (1997), 6760-6774
doi:10.1103/PhysRevD.55.6760
[arXiv:hep-ph/9703464 [hep-ph]].

\bibitem{Horava:2009uw}
P.~Horava,
Phys. Rev. D \textbf{79} (2009), 084008
doi:10.1103/PhysRevD.79.084008
[arXiv:0901.3775 [hep-th]].

\bibitem{Carroll:2001ws}
S.~M.~Carroll, J.~A.~Harvey, V.~A.~Kostelecky, C.~D.~Lane and T.~Okamoto,
Phys. Rev. Lett. \textbf{87} (2001), 141601
doi:10.1103/PhysRevLett.87.141601
[arXiv:hep-th/0105082 [hep-th]].


\bibitem{Kostelecky:1988zi}
V.~A.~Kostelecky and S.~Samuel,
Phys. Rev. D \textbf{39} (1989), 683
doi:10.1103/PhysRevD.39.683


\bibitem{Bluhm:2008yt}
R.~Bluhm, N.~L.~Gagne, R.~Potting and A.~Vrublevskis,
Phys. Rev. D \textbf{77} (2008), 125007
[erratum: Phys. Rev. D \textbf{79} (2009), 029902]
doi:10.1103/PhysRevD.79.029902
[arXiv:0802.4071 [hep-th]].

\bibitem{Kostelecky:1989jw}
V.~A.~Kostelecky and S.~Samuel,
Phys. Rev. D \textbf{40} (1989), 1886-1903
doi:10.1103/PhysRevD.40.1886

\bibitem{Bertolami:2005bh}
O.~Bertolami and J.~Paramos,
Phys. Rev. D \textbf{72} (2005), 044001
doi:10.1103/PhysRevD.72.044001
[arXiv:hep-th/0504215 [hep-th]].

\bibitem{Casana:2017jkc}
R.~Casana, A.~Cavalcante, F.~P.~Poulis and E.~B.~Santos,
Phys. Rev. D \textbf{97} (2018) no.10, 104001
doi:10.1103/PhysRevD.97.104001
[arXiv:1711.02273 [gr-qc]].

\bibitem{Ovgun:2018ran}
A.~Ovg{\"u}n, K.~Jusufi and I.~Sakalli,
Annals Phys. \textbf{399} (2018), 193-203
doi:10.1016/j.aop.2018.10.012
[arXiv:1805.09431 [gr-qc]].

\bibitem{Yang:2018zef}
R.~J.~Yang, H.~Gao, Y.~Zheng and Q.~Wu,
Commun. Theor. Phys. \textbf{71} (2019) no.5, 568-572
doi:10.1088/0253-6102/71/5/568
[arXiv:1809.00605 [gr-qc]].

\bibitem{Li:2020dln}
Z.~Li and A.~{\"O}vg{\"u}n,
Phys. Rev. D \textbf{101} (2020) no.2, 024040
doi:10.1103/PhysRevD.101.024040
[arXiv:2001.02074 [gr-qc]].

\bibitem{DCarvalho:2021zpf}
{\'I}.~D.~D.Carvalho, G.~Alencar, W.~M.~Mendes and R.~R.~Landim,
EPL \textbf{134} (2021) no.5, 51001
doi:10.1209/0295-5075/134/51001
[arXiv:2103.03845 [gr-qc]].

\bibitem{Jha:2020pvk}
S.~K.~Jha and A.~Rahaman,
Eur. Phys. J. C \textbf{81} (2021) no.4, 345
doi:10.1140/epjc/s10052-021-09132-6
[arXiv:2011.14916 [gr-qc]].

\bibitem{Oliveira:2021abg}
R.~Oliveira, D.~M.~Dantas and C.~A.~S.~Almeida,
EPL \textbf{135} (2021) no.1, 10003
doi:10.1209/0295-5075/ac130c
[arXiv:2105.07956 [gr-qc]].

\bibitem{Xu:2022frb}
R.~Xu, D.~Liang and L.~Shao,
Phys. Rev. D \textbf{107} (2023) no.2, 024011
doi:10.1103/PhysRevD.107.024011
[arXiv:2209.02209 [gr-qc]].

\bibitem{Filho:2022yrk}
A.~A.~A.~Filho, J.~R.~Nascimento, A.~Y.~Petrov and P.~J.~Porf{\'\i}rio,
Phys. Rev. D \textbf{108} (2023) no.8, 085010
doi:10.1103/PhysRevD.108.085010
[arXiv:2211.11821 [gr-qc]].

\bibitem{Uniyal:2022xnq}
A.~Uniyal, S.~Kanzi and {\.I}.~Sakall{\i},
Eur. Phys. J. C \textbf{83} (2023) no.7, 668
doi:10.1140/epjc/s10052-023-11846-8
[arXiv:2207.10122 [hep-th]].

\bibitem{Liu:2022dcn}
W.~Liu, X.~Fang, J.~Jing and J.~Wang,
Eur. Phys. J. C \textbf{83} (2023) no.1, 83
doi:10.1140/epjc/s10052-023-11231-5
[arXiv:2211.03156 [gr-qc]].

\bibitem{AraujoFilho:2024ykw}
A.~A.~Ara{\'u}jo Filho, J.~R.~Nascimento, A.~Y.~Petrov and P.~J.~Porf{\'\i}rio,
JCAP \textbf{07} (2024), 004
doi:10.1088/1475-7516/2024/07/004
[arXiv:2402.13014 [gr-qc]].



\bibitem{Remillard:2006fc}
R.~A.~Remillard and J.~E.~McClintock,
Ann. Rev. Astron. Astrophys. \textbf{44} (2006), 49-92
doi:10.1146/annurev.astro.44.051905.092532
[arXiv:astro-ph/0606352 [astro-ph]].

\bibitem{Stella:1998mq}
L.~Stella and M.~Vietri,
Phys. Rev. Lett. \textbf{82} (1999), 17-20
doi:10.1103/PhysRevLett.82.17
[arXiv:astro-ph/9812124 [astro-ph]].

\bibitem{Stella:1997tc}
L.~Stella and M.~Vietri,
Astrophys. J. Lett. \textbf{492} (1998), L59
doi:10.1086/311075
[arXiv:astro-ph/9709085 [astro-ph]].

\bibitem{Cadez:2008iv}
A.~Cadez, M.~Calvani and U.~Kostic,
Astron. Astrophys. \textbf{487} (2008), 527-532
doi:10.1051/0004-6361:200809483
[arXiv:0809.1783 [astro-ph]].

\bibitem{Kostic:2009hp}
U.~Kostic, A.~Cadez, M.~Calvani and A.~Gomboc,
Astron. Astrophys. \textbf{496} (2009), 307
doi:10.1051/0004-6361/200811059
[arXiv:0901.3447 [astro-ph.HE]].

\bibitem{Abramowicz:2003xy}
M.~A.~Abramowicz, V.~Karas, W.~Kluzniak, W.~H.~Lee and P.~Rebusco,
Publ. Astron. Soc. Jap. \textbf{55} (2003), 466-467
doi:10.1093/pasj/55.2.467
[arXiv:astro-ph/0302183 [astro-ph]].

\bibitem{Kluzniak:2002bb}
W.~Kluzniak and M.~A.~Abramowicz,
[arXiv:astro-ph/0203314 [astro-ph]].

\bibitem{Nowak:1996hg}
M.~A.~Nowak, R.~V.~Wagoner, M.~C.~Begelman and D.~E.~Lehr,
Astrophys. J. Lett. \textbf{477} (1997), L91
doi:10.1086/310534
[arXiv:astro-ph/9612142 [astro-ph]].

\bibitem{Torok:2010rk}
G.~Torok, P.~Bakala, E.~Sramkova, Z.~Stuchlik and M.~Urbanec,
Astrophys. J. \textbf{714} (2010), 748-757
doi:10.1088/0004-637X/714/1/748
[arXiv:1008.0088 [astro-ph.HE]].

\bibitem{Kotrlova:2020pqy}
A.~Kotrlov{\'a}, E.~{\v{S}}r{\'a}mkov{\'a}, G.~T{\"o}r{\"o}k, K.~Goluchov{\'a}, J.~Hor{\'a}k, O.~Straub, D.~Lancov{\'a}, Z.~Stuchl{\'\i}k and M.~A.~Abramowicz,
Astron. Astrophys. \textbf{643} (2020), A31
doi:10.1051/0004-6361/201937097
[arXiv:2008.12963 [astro-ph.HE]].

\bibitem{Stella:1999sj}
L.~Stella, M.~Vietri and S.~Morsink,
Astrophys. J. Lett. \textbf{524} (1999), L63-L66
doi:10.1086/312291
[arXiv:astro-ph/9907346 [astro-ph]].


\bibitem{Motta:2013wwa}
S.~E.~Motta, T.~Mu{\~n}oz-Darias, A.~Sanna, R.~Fender, T.~Belloni and L.~Stella,
Mon. Not. Roy. Astron. Soc. \textbf{439} (2014), 65
doi:10.1093/mnrasl/slt181
[arXiv:1312.3114 [astro-ph.HE]].

\bibitem{Ingram:2014ara}
A.~Ingram and S.~Motta,
Mon. Not. Roy. Astron. Soc. \textbf{444} (2014) no.3, 2065-2070
doi:10.1093/mnras/stu1585
[arXiv:1408.0884 [astro-ph.HE]].

\bibitem{Maselli:2014fca}
A.~Maselli, L.~Gualtieri, P.~Pani, L.~Stella and V.~Ferrari,
Astrophys. J. \textbf{801} (2015) no.2, 115
doi:10.1088/0004-637X/801/2/115
[arXiv:1412.3473 [astro-ph.HE]].

\bibitem{Ahmed:2026hmy}
F.~Ahmed, A.~Al-Badawi, S.~Murodov, B.~Rahmatov and J.~Rayimbaev,
[arXiv:2607.01723 [gr-qc]].

\bibitem{Jumaniyozov:2025dyy}
S.~Jumaniyozov, S.~Murodov, J.~Rayimbaev, I.~Ibragimov, B.~Madaminov, S.~Urinbaev and A.~Abdujabbarov,
Eur. Phys. J. C \textbf{85} (2025) no.7, 797
doi:10.1140/epjc/s10052-025-14522-1


\bibitem{Jumaniyozov:2025wcs}
S.~Jumaniyozov, M.~Zahid, M.~Alloqulov, I.~Ibragimov, J.~Rayimbaev and S.~Murodov,
Eur. Phys. J. C \textbf{85} (2025) no.2, 126
doi:10.1140/epjc/s10052-025-13863-1

\bibitem{Jumaniyozov:2024eah}
S.~Jumaniyozov, S.~U.~Khan, J.~Rayimbaev, A.~Abdujabbarov, S.~Urinbaev and S.~Murodov,
Eur. Phys. J. C \textbf{84} (2024) no.9, 964
doi:10.1140/epjc/s10052-024-13351-y


\bibitem{Wang:2021gtd}
Z.~Wang, S.~Chen and J.~Jing,
Eur. Phys. J. C \textbf{82} (2022) no.6, 528
doi:10.1140/epjc/s10052-022-10475-x
[arXiv:2112.02895 [gr-qc]].

\bibitem{Zhang:2025acq}
H.~Y.~Zhang, Y.~P.~Hu and Y.~S.~An,
Eur. Phys. J. C \textbf{85} (2025) no.7, 725
doi:10.1140/epjc/s10052-025-14448-8
[arXiv:2503.02323 [gr-qc]].

\bibitem{Mustafa:2024mvx}
G.~Mustafa, S.~K.~Maurya, P.~Channuie, A.~Bouzenada, A.~Abd-Elmonem and N.~Alhubieshi,
Phys. Dark Univ. \textbf{47} (2025), 101753
doi:10.1016/j.dark.2024.101753


\bibitem{Liu:2025oho}
J.~Z.~Liu, S.~P.~Wu, S.~W.~Wei and Y.~X.~Liu,
Sci. China Phys. Mech. Astron. \textbf{69} (2026) no.7, 270411
doi:10.1007/s11433-026-2961-8
[arXiv:2510.16731 [gr-qc]].



\bibitem{Liu:2024axg}
J.~Z.~Liu, W.~D.~Guo, S.~W.~Wei and Y.~X.~Liu,
Eur. Phys. J. C \textbf{85} (2025) no.2, 145
doi:10.1140/epjc/s10052-025-13859-x
[arXiv:2407.08396 [gr-qc]].

\bibitem{Kluzniak:2013}
W.~Klu{\'z}niak and D.~Rosi{\'n}ska,
Mon. Not. Roy. Astron. Soc. \textbf{434} (2013) no.4, 2825--2829
doi:10.1093/mnras/stt1185.

\bibitem{Rayimbaev:2021kjs}
J.~Rayimbaev, S.~Shaymatov and M.~Jamil,
Eur. Phys. J. C \textbf{81} (2021) no.8, 699
doi:10.1140/epjc/s10052-021-09488-9
[arXiv:2107.13436 [gr-qc]].

\bibitem{Stuchlik:2022xtq}
Z.~Stuchl{\'\i}k and J.~Vrba,
Astrophys. J. \textbf{935} (2022) no.2, 91
doi:10.3847/1538-4357/ac7f27
[arXiv:2208.02612 [gr-qc]].

\bibitem{Motta:2013wga}
S.~E.~Motta, T.~M.~Belloni, L.~Stella, T.~Mu{\~n}oz-Darias and R.~Fender,
Mon. Not. Roy. Astron. Soc. \textbf{437} (2014) no.3, 2554-2565
doi:10.1093/mnras/stt2068
[arXiv:1309.3652 [astro-ph.HE]].



\bibitem{QiQi:2026pnb}
Q.~Qi, Y.~Sang and X.~M.~Kuang,
Sci. China Phys. Mech. Astron. \textbf{69} (2026) no.6, 260414
doi:10.1007/s11433-026-2960-1
[arXiv:2601.06491 [gr-qc]].

\bibitem{Foreman-Mackey:2012any}
D.~Foreman-Mackey, D.~W.~Hogg, D.~Lang and J.~Goodman,
Publ. Astron. Soc. Pac. \textbf{125} (2013), 306-312
doi:10.1086/670067
[arXiv:1202.3665 [astro-ph.IM]].

\bibitem{Ali:2026pmh}
R.~H.~Ali, M.~H.~Wu, H.~Guo and X.~M.~Kuang,
Eur. Phys. J. C \textbf{86} (2026) no.5, 540
doi:10.1140/epjc/s10052-026-15772-3
[arXiv:2602.11525 [gr-qc]].

\bibitem{Orosz:2011ki}
J.~A.~Orosz, J.~F.~Steiner, J.~E.~McClintock, M.~A.~P.~Torres, R.~A.~Remillard, C.~D.~Bailyn and J.~M.~Miller,
Astrophys. J. \textbf{730} (2011), 75
doi:10.1088/0004-637X/730/2/75
[arXiv:1101.2499 [astro-ph.SR]].

\bibitem{Remillard:2002cy}
R.~A.~Remillard, M.~P.~Muno, J.~E.~McClintock and J.~A.~Orosz,
Astrophys. J. \textbf{580} (2002), 1030-1042
doi:10.1086/343791
[arXiv:astro-ph/0202305 [astro-ph]].

\bibitem{Reid:2014ywa}
M.~J.~Reid, J.~E.~McClintock, J.~F.~Steiner, D.~Steeghs, R.~A.~Remillard, V.~Dhawan and R.~Narayan,
Astrophys. J. \textbf{796} (2014), 2
doi:10.1088/0004-637X/796/1/2
[arXiv:1409.2453 [astro-ph.GA]].

\bibitem{Liu:2023vfh}
C.~Liu, H.~Siew, T.~Zhu, Q.~Wu, Y.~Sun, Y.~Zhao and H.~Xu,
JCAP \textbf{11} (2023), 096
doi:10.1088/1475-7516/2023/11/096
[arXiv:2305.12323 [gr-qc]].

\bibitem{Sharma:2017wfu}
S.~Sharma,
Ann. Rev. Astron. Astrophys. \textbf{55} (2017), 213-259
doi:10.1146/annurev-astro-082214-122339
[arXiv:1706.01629 [astro-ph.IM]].

\bibitem{Zhang:2025acq}
H.~Y.~Zhang, Y.~P.~Hu and Y.~S.~An,
Eur. Phys. J. C \textbf{85} (2025) no.7, 725
doi:10.1140/epjc/s10052-025-14448-8
[arXiv:2503.02323 [gr-qc]].


\end{thebibliography}

\end{document}